\documentclass[structabstract]{aa}  
\usepackage{graphicx}
\usepackage{txfonts}
\usepackage{longtable}

\usepackage{natbib}

\def\llm{{\sc LLmodels}}

\def\width9{{\sc WIDTH9}}

\def\logg{\log(g)}
\def\teff{T_{\rm eff}}

\def\hbeta{\mathrm{H}\beta}
\def\hgamma{\mathrm{H}\gamma}
\def\ddafit{{\sc DDAFit}}
\def\synth3{{\sc Synth3}}

\def\synthmag{{\sc Synthmag}}

\def\Lsun{L_{\odot}}
\def\Rsun{R_{\odot}}
\def\Msun{M_{\odot}}

\def\aql{$10$~Aql}

\begin{document}
   \title{A self consistent chemically stratified atmosphere model for the roAp star 10 Aquilae}

   \subtitle{}

   \author{N. Nesvacil\inst{1,2}
          \and
	  D. Shulyak\inst{3}
          \and
	  T. A. Ryabchikova\inst{4}
	  \and
	  O. Kochukhov\inst{5}
          \and
	  A. Akberov\inst{4}
	  \and
	  W. Weiss\inst{2}}
          
\offprints{N. Nesvacil \\
\email{nicole.nesvacil@univie.ac.at}}

   \institute{Department of Radiotherapy, Medical University of Vienna, Waehringer Guertel 18-20, 1090 Vienna \and
   	      Institute for Astronomy (IfA), University of Vienna,
              Tuerkenschanzstrasse 17, A-1180 Vienna\and
       	      Institute of Astrophysics, Georg-August University, Friedrich-Hund-Platz 1, D-37077 G\"ottingen, Germany\and
	      Institute of Astronomy, Russian Academy of Sciences, Pyatnitskaya 48, 109017 Moscow, Russia\and
	      Department of Physics and Astronomy, Uppsala University, Box 516, 75120 Uppsala, Sweden             }

   \date{Received / Accepted}

  \abstract
   {Chemically peculiar A type (Ap) stars are a subgroup of the CP2 stars which exhibit anomalous overabundances of numerous elements, e.g. Fe,
   Cr, Sr and rare earth elements. The pulsating subgroup of the Ap stars, the roAp stars, present ideal laboratories to observe and model
   pulsational signatures as well as the interplay of the pulsations with strong magnetic fields and vertical abundance gradients.}
   {Based on high resolution spectroscopic observations and observed stellar energy distributions
   we construct a self consistent model atmosphere, that accounts for modulations of the temperature-pressure structure 
   caused by vertical abundance gradients, for the roAp star 10 Aquilae (HD~176232). 
   We demonstrate that such an analysis can be used to determine precisely the fundamental atmospheric parameters required for
   pulsation modelling.}
   {Average abundances were derived for 56 species. For  Mg, Si, Ca, Cr, Fe, Co, Sr, Pr, and Nd vertical stratification profiles were empirically derived using the 
   \ddafit\ minimization routine together with the magnetic spectrum synthesis code \synthmag. 
   Model atmospheres were computed with the \llm\ code which accounts
   for the individual abundances and stratification of chemical elements.}
   {For the final model atmosphere $\teff=7550$ K and $\logg=3.8$ were adopted. While Mg, Si, Co and Cr exhibit steep abundance gradients 
   Ca, Fe and Sr showed much wider abundance gradients between log$\tau_{\rm5000}=-1.5$
   and $0.5$. Elements Mg and Co were found to be the least stratified, while Ca and Sr showed strong depth variations in abundance of up to $\approx 6$~dex.}
  {}


   \maketitle
%

\section{Introduction}

10\,Aql (HR\,7167, HD\,176232, HIP\,93179) is one of the brightest \textit{rapidly oscillating Ap} (roAp) stars. These stars exhibit high-overtone, low-degree,
non-radial $p$-mode pulsations with periods in the range of 6--24 minutes \citep{KM00,2012MNRAS.421L..82AAKR2012}. The
most distinquished characteristic of roAp pulsations is the dependence of the radial velocity (RV) amplitude and phase on the particular chemical
element/ion. After discovery of stratified abundance distribution in the atmospheres of Ap stars it became possible to connect the observed
pulsational characteristics with the line depth formation in stratified atmospheres \citep{TR2002}. However, to get a self-consistent model
atmosphere of an Ap star one needs to take into account anomalous abundances of most elements of the Mendelejev table as well as the 
stratified distribution of those elements that provide significant contribution to the continuum and line opacities. Recently, such modelling
was performed for a few roAp stars: $\alpha$~Cir \citep{2009A&A...499..851K}, HD~24712 \citep{2009A&A...499..879S}, and HD~101065 - Przybylski' star
\citep{2010A&A...520A..88S}. \citet{2010MNRAS.403.1729S} showed that the use of the self-consistent model atmopshere of HD~24712 allowed to provide
better modelling of pulsations and to derive the theoretical distribution of RV amplitudes and phases over the stellar atmosphere, compatible 
with the observed RV pulsational characteristics. 

A first abundance analysis based on high resolution spectra was performed by \citet{Ryabchikova2000}. The mean magnetic field
modulus 1.2 kG was determined by comparison with magnetic spectrum synthesis calculations by \citet{Kochukhov2002}.
Pulsations in 10\,Aql were discovered in a search for stars in the northern hemisphere by \citet{Heller1988}. Pulsations in spectral lines were detected by \citet{Kochukhov2002}.
The star exhibits very sharp spectral lines which is typical for very slow rotators. Being one of the brightest roAp stars, 10\,Aql was an excellent target for 
a photometric
campaign with the space telescope MOST \citep{Huber2008}. From their detailed analysis Huber et al. could derive a lower limit for the rotation period, 
$P_{rot}>$ 1 month, which was later confirmed by the study of RV pulsations \citep{2008MNRAS.389..903S}. The authors performed detailed measurements of RV
pulsations in the atmosphere of 10\,Aql and found a phase jump in the RV pulsations derived from lines of the rare-earth (REE) elements Nd and Pr. According to
theoretical predictions \citep{Khomenko2009,Sousa2011} the phase jumps may be associated with the so-called nodal regions. To get a proper pulsation
model for any roAp star one needs to derive a depth-dependence of the pulsational characteristics. 

The main goal of the present paper is to obtain the chemical and atmospheric structure of 10\,Aql and global parameters such as radius, mass and luminosity 
based on the available high quality spectroscopic, spectrophotometric and photometric observations.    
The paper is organized as follows: spectroscopic observations and data reduction are shortly presented in Sect.~\ref{sec:obs}, the methods of the analysis
are given in Sect.~\ref{sec:meth}. The resulting abundances, element stratification and atmospheric structure calculations are presented in
Sect.~\ref{sec:res}. In Sect.~\ref{sec:concl} we discuss the importance of the detailed abundance and stratification analysis for atmospheric structure
modelling.

\section{Observations}
\label{sec:obs}

As 10\,Aql is a southern hemisphere object, the star was included in a larger survey for high resolution spectra of 
sharp-lined Ap stars with the ESO-VLT UVES spectrograph, 68.D-0254(A) 
(PI Kochukhov). 
Dichroic standard settings DIC$\# 2$ 
centered at 346+580 nm and 437+860 nm were used to obtain a wavelength coverage from 3030 to 10\,400 \AA \, with only a few minor gaps. The spectral resolution
resulting from the chosen slit width of 0.5 arcseconds is $R \approx$ 80\,000 with a signal-to-noise ratio of $450-550$. 
Data reduction of UVES observations \citep[e.g.][]{RyabchikovaCaIso} was performed with the ESO UVES pipeline \citep{UVESpipe}. 
Additional continuum rectification was performed using a dedicated IDL programme. 

\section{Methods}
\label{sec:meth}
\subsection{Abundance Analysis}
As a first step in this investigation, a detailed abundance analysis was made for \aql\ in order to derive input parameters for the first calculation 
of an individual model atmosphere with the \llm\ code by \citet{LLModels}. This code takes into account individual abundances 
for all elements and can additionally accept vertical distributions of chemical elements as input. Abundances were derived from equivalent width measurements of 490 lines
obtained with the IDL programme {\it ROTATE}, a graphical interface for comparison of observed and synthetic spectra, and a modified version 
of the {\it WIDTH9} code 
\citep{KuruczCD18,1996ASPC..108..198T}, which includes additional magnetic line broadening. Abundances for 56 species were calculated. 
Atomic line data were extracted from the Vienna Atomic Line Database (VALD II, \citet{VALD_2,VALD_3,VALD_4}). 
For this initial abundance analysis atmospheric parameters from \citet{Ryabchikova2000} were used, i.e. $\teff$= 7650 K, $\log g$= 4.0 and
a mean magnetic field modulus of $<B>$=1.2 kG. 
The results were then used to compute an individual model atmosphere. 
Comparison of observed H$\alpha$ line profiles with the synthetic profiles based on this new model
revealed the necessity to reduce the effective temperature by 100~K in order to obtain a better fit. 
Table \ref{abundancesI} lists the resulting abundances of chemical elements.
These abundances were then used to identify elements that appear to be vertically stratified in the atmosphere of the star. 
Elements which revealed systematic differences between abundances derived from lines of different ionization stages 
are expected to be inhomogeneously distributed in the stellar atmosphere.

\subsection{Stratification Analysis}
Vertical abundance stratification is considered to be responsible for an impossibility to model cores and wings of strong lines with developed Stark wings with the same 
abundance, or as an impossibility to reproduce lines belonging to different ionization stages of the same element or high- and low-excitation lines with a chemically homogeneous
atmosphere \citep{TRUps2003}. All three effects were detected during
the initial abundance analysis and nine elements were included in subsequent further stratification analysis.

Starting from the $\teff=7550$~K, $\logg=4.0$ model atmosphere with a peculiar but vertically homogeneous chemical composition, 
stratification profiles of Mg, Si, Ca, Cr, Fe, Co, and Sr were computed with the \ddafit\ IDL-based code \citep{OlegCodes}. 
This code iteratively fits a vertical abundance 
distribution to an unlimited set of observed spectral lines. The stratification profiles are defined by four parameters: 
upper atmospheric abundance, abundance in deep atmospheric layers, the vertical position of the abundance jump,
and the width of the transition region where abundances change
between the two values. All four parameters are optimized simultaneously for one element at a time, but a fixed 
stratification of other elements can also be taken into account.

Element Co was included in the model calculations only after a first consistent stratified model had been derived 
by \citet{Thesis}, as hyperfine structure data, which are necessary to
synthesize observed Co line profiles, were not available at the beginning of our study. 
Stratification of rare earth elements was derived using a semi-empirical fitting procedure based on observed 
equivalent widths under NLTE assumption. Therefore the shape of derived abundance profiles may deviate from a single-step function.
The method was described in \citet{2009A&A...495..297M} and \citet{2005A&A...441..309M}.

\subsubsection{Line Selection}
During this first stratification analysis 
large numbers of spectral lines were inspected most carefully in order to assure the employment of the best suitable lines 
with the most accurate atomic 
parameters available. For the stratification analysis strong and weak lines of different ions are used to sample a wide range 
of atmospheric layers. 
Lines showing unusually broadened wings and relatively shallow cores, which were suspected to be a result of vertical 
element stratification, were also selected.  

A list of the atomic parameters used for further analysis is given for each element in Table~\ref{atomicparsALL} 
in the Online material.
Atomic line parameters were extracted from VALD using the \textit{long format} option for magnetic spectrum synthesis. 
The original sources of atomic data included in VALD are given in the corresponding table caption.

 
Part of the line sets selected for stratification analysis in this work correspond to the ones presented by
\citet{OK2006HD133792} and \citet{RyabchikovaCaIso}. The latter have studied simultaneously the effects of 
peculiar Ca isotopic mixtures on 
line shapes and stratification profiles in a large number of Ap stars. In the present investigation the same Ca lines 
were used for the analysis, with the exception of 
lines which would be sensitive to the isotopic mixture. Therefore, ignoring this effect does not have a negative influence 
on the results of the present study.

In order to estimate which atmospheric layers are sampled by the selected set of lines, formation depths of each line 
were calculated based on the method proposed
by \citet{Achmad1991}, which was implemented in our spectrum synthesis code. Based on the calculated contribution functions, 
a range of atmospheric depths which contribute most to the opacities at given wavelength points was derived for each line profile used in our stratification 
analysis. Figure \ref{CFFeSr} shows how these so called line formation depths 
change when Fe or Sr stratification
are introduced in model atmosphere calculations. In the case of Fe a few lines were available in the high energy region 
to sample the upper atmosphere layers and constrain the
upper abundance value. High excitation Fe lines are indicated to be useful for sampling lower atmospheric layers. 
For comparison, a similar plot is shown for a typical 
Sr stratification. The available observed Sr lines appear to be sensitive only to a narrow atmosphere depth region 
in the homogeneous as well as in the
stratified case. Therefore the derived location of the jump and the deeper atmospheric abundance are considered to be more reliable than the upper atmospheric abundance, 
which is determined by the depth of the line cores of the strongest lines. 

For iron, a large number of observed spectral lines was available. Lines from UV, visual and near infrared wavelength 
regions were selected to ensure a better
sampling of the atmospheric layers where the abundance jump occurs, and put better constraints on the upper abundance value.   
In order to test the effect of the choice of line list on the resulting stratification profile two Fe distributions 
were calculated from a complete line list and one
excluding UV and infrared lines. The difference between lower atmospheric abundances, and locations and slopes of the abundance 
jumps was found to be small, i.e. within the formal
errors of the fitting routine. The upper atmospheric abundance derived from the line list including strong UV lines
was found to be 1.7~dex lower than the one derived
from the visual lines only. However, as only two lines in the sample are really sensitive to these layers, 
the uncertainty of the derived abundance is expected to be of the order of 1~dex.

In order to obtain the best possible sampling of atmosphere layers for all elements, the final stratification line lists 
aimed to include as many high excitation, 
UV and infrared lines as possible, in addition to weak and strong lines from the visual range and lines that exhibit 
unusually broad wings and shallow cores due to vertical stratification. 

\begin{figure*}
\includegraphics[angle=270, width=0.5\textwidth]{CFFe_2panels.ps}
\includegraphics[angle=270, width=0.5\textwidth]{CFSr_2panels.ps}
\caption{Range of atmosphere depths that contribute most to formation of selected line profiles, based on contribution functions according to \citet{Achmad1991} for Fe (left column) and Sr (right column). Upper panels show example 
stratification profile and formation depth ranges of selected lines in stratified model atmosphere, lower panels show formation depth ranges for lines in homogeneous model atmosphere.}
\label{CFFeSr}
\end{figure*}

\subsubsection{Influence of calculation setup on profile fitting procedure}

For several cases we find that the fitting routine 
\ddafit\ produces results with large nominal errors for
abundances or width of the abundance jump, especially when the atmospheric layers are probed by only a few spectral lines.
Therefore small changes in the model atmosphere setup could have a large effect 
on the resulting parameters, which might be
misinterpreted as being caused by real chemical gradients. 
A few tests were performed to investigate the behaviour of the code in more detail. 

In one experiment, a set of Fe stratification profiles was calculated based on the same model atmosphere 
but with different depth scales used to compute numerical derivatives in the minimization part of the code.
Here we used optical depth $\log\tau_{\rm 5000}$ or column mass $M$.
The resulting profiles showed no difference between abundance values in deep atmosphere layers larger than 0.1~dex, 
and location as well as slope of the transition region were comparable. 
The upper atmospheric abundance
differed by 0.5~dex. This difference could be removed by using a model atmosphere with 288 layers instead of 72 
in \ddafit\ calculations in log$\tau_{\rm 5000}$. 
As the increased number of layers also increases overall calculation times by up to a few hours, 
for the rest of the analysis column mass was used as a depth scale.
Finally, results obtained this way are plotted on an optical depth scale. In the case of relatively wide jumps, 
this transformation results in rather smoothed shapes of the distribution profile instead of a simple two-step profile, 
due to the non-linear relation between $\tau_{\rm 5000}$ and $M$. 

In another experiment, we computed different stratification profiles for seven elements using two model atmospheres 
with the same temperature, gravity
and chemical composition, but using 80 or 118 layers. The resulting profiles were qualitatively identical for most elements 
except for Ca, Fe and Sr. In those three 
cases the model with higher depth resolution resulted in smaller abundances (-5, -1.5, and -2.5~dex) for upper atmospheric layers,
while the overall distributions remained the same as for the model sampled on 80 depths. 
This finding indicates that for these three elements upper abundances are only constrained by a few sensitive lines and small
changes in atmosphere depth resolution might therefore have a large impact on the upper atmospheric abundances derived with \ddafit. 
Adjusting the derivative step
size used in \ddafit\ decreased the formal error of the fitting procedure for the upper atmospheric abundance value 
below 0.8~dex. The difference of upper atmospheric 
Ca and Fe abundances between the two models thereby decreased as well from -1.4 to -0.7~dex, while it remained $\approx-2.5$~dex 
for Sr. This result indicates that 
the upper abundance of Sr is indeed not very well constrained by the available set of spectral lines. 
Abundance values in higher atmospheric layers are very sensitive to small
changes in model atmosphere and calculation setups for all three elements. Other fitting parameters for Ca, Fe and Sr, and overall 
stratification profiles of all other elements seem, however, not very much affected by such changes. 

During the iterative calculation of self consistent stratification models it is sometimes necessary to adjust $\teff$ 
to obtain better fits to observed fluxes.
Similar to what was described above another test showed that a change of $\pm$100~K can result in a change of the upper atmospheric 
abundance 
by up to a few dex, if most available spectral lines are not very sensitive in these layers. 
Location and slope of the abundance gradient, as well as lower atmospheric abundances
were not affected noticeably by such changes in $\teff$. 

\subsubsection{Iterative Stratification Calculation}
\label{sec:isc}

The basic steps of the iterative procedure of atmospheric parameters and abundance determination
were outlined in \citet{2009A&A...499..879S}.
After the first set of stratification profiles had been obtained, new model atmospheres for a set of $[\teff,\logg]$ pairs
were calculated using the derived stratification as input. 
For testing the quality of various atmosphere models observed hydrogen line profiles and spectral energy distributions were used. 
In particular we made use of spectrophotometric observations by \citet{1989A&AS...81..221A} and
\citet{1976ApJS...32....7B}, ultraviolet energy distributions from the 
IUE satellite mission\footnote{http://archive.stsci.edu/iue/}, 
and spectrophotometric observations obtained by the STIS instrument\footnote{http://www.stsci.edu/hst/stis}
mounted at the Hubble Space Telescope. 
A convergence of the stratification models was reached, when more iterations would not have improved the fit between 
synthetic line profiles and other observables any further,
taking into account the formal error on abundances and step parameters of the \ddafit\ procedure. 

\section{Results}
\label{sec:res}
\subsection{Abundance analysis}

Mean abundances were computed for a converged stratified model and are summarized in the last three columns of Table 
\ref{abundancesI}. For comparison, the results of the abundance analysis 
based on
the initial homogeneous model ($\teff=7550$ K and $\logg=4.0$) and an intermediate stratified model with lower temperature
are listed (columns 2 and 3 respectively).
Including stratification in the model atmosphere calculation did not affect the overall abundances of non-stratified elements. 
The difference between homogeneous and 
stratified models stayed within 0.1~dex for most elements, which is well within the accuracy of the method of 
abundance determination by fitting equivalent widths.
The comparison of abundances from two stratified models with different temperatures illustrates that the mean abundances 
of homogeneously distributed elements were 
not affected by the changes in model structure between different iteration steps of the stratification fitting procedure. 

\begin{center}
\begin{table}
\caption[]{Mean abundances and standard deviations derived from observed EQW with the starting model, 
an intermediate model with $\teff=7450$ K and $\logg=4.0$, 
and the final model with $\teff=7550$ K and $\logg=3.8$. 
Column "$\#$" denotes the number of lines measured for each ion. Stratified elements are marked by asterisks.}
\label{abundancesI}
\begin{scriptsize}
\begin{tabular}{l|cc|cc|cc|c}
\hline\hline
ion& log$N/N_{\rm tot}$ & $\sigma$ &log$N/N_{\rm tot}$ &$\sigma$ &log$N/N_{\rm tot}$ &$\sigma$ &$\#$  \\
\hline  					     
\ion{C}{i}     &-4.21	        &	 0.06   &   -4.28   &0.06 &  	-4.24	&0.06 &  3     \\ 
\ion{O}{i}	&-3.89  	&	 0.38	&   -3.96   &0.40 &  	-3.92	&0.38 &  9     \\ 
\ion{Na}{i}	&-6.21  	&	 0.10	&   -6.31   &0.11 &  	-6.29	&0.12 &  2     \\ 
\ion{Mg}{i}	&-4.60  	&	 0.12	&   $\ast$    &     &  	$\ast$	&     &  2     \\ 
\ion{Mg}{ii}	&-4.39  	&		&   $\ast$    &	  &  	$\ast$	&     &  1     \\ 
\ion{Al}{i}	&-5.84  	&	 0.03	&   -5.93   &0.03 &  	-5.90	&0.03 &  2     \\ 
\ion{Al}{ii}	&-4.90  	&		&   -4.95   &	  &  	-4.91	&     &  1     \\ 
\ion{Si}{i}	&-4.49  	&	 0.54	&   $\ast$    &     &  	$\ast$	&     &  7     \\ 
\ion{S}{i}	&-5.21  	&	 0.14	&   -5.29   &0.14 &  	-5.27	&0.14 &  3     \\ 
\ion{Ca}{i}	&-5.49  	&	 0.29	&   $\ast$    &     &  	$\ast$	&     &  4     \\ 
\ion{Ca}{ii}	&-4.67  	&	 0.03	&   $\ast$    &     &  	$\ast$	&     &  2     \\ 
\ion{Sc}{ii}	&-9.60  	&	 0.27	&   -9.76   &0.27 &  	-9.76	&0.27 &  6     \\ 
\ion{Ti}{i}	&-7.26  	&	 0.34	&   -7.37   &0.35 &  	-7.37	&0.35 &  12    \\ 
\ion{Ti}{ii}	&-7.31  	&	 0.19	&   -7.37   &0.18 &  	-7.36	&0.19 &  19    \\ 
\ion{V}{ii}	&-7.65  	&	 0.40	&   -7.80   &0.40 &  	-7.80	&0.40 &  4     \\ 
\ion{Cr}{i}	&-5.28  	&	 0.21	&   $\ast$    &     &  	$\ast$	&     &  33    \\ 
\ion{Cr}{ii}	&-5.38  	&	 0.28	&   $\ast$    &     &  	$\ast$	&     &  49    \\ 
\ion{Mn}{i}	&-6.35  	&	 0.30	&   -6.45   &0.30 &  	-6.44	&0.30 &  7     \\ 
\ion{Mn}{ii}	&-6.42  	&	 0.02	&   -6.51   &0.08 &  	-6.50	&0.09 &  2     \\ 
\ion{Fe}{i}	&-4.45  	&	 0.34	&   $\ast$    &     &  	$\ast$	&     &  105   \\ 
\ion{Fe}{ii}	&-4.09  	&	 0.36	&   $\ast$    &     &  	$\ast$	&     &  48    \\ 
\ion{Co}{i}	&-5.95  	&	 0.19	&   $\ast$    &     &  	$\ast$	&     &  19    \\ 
\ion{Co}{ii}	&-5.72  	&	 0.10	&   $\ast$    &     &  	$\ast$	&     &  2     \\ 
\ion{Ni}{i}	&-6.47  	&	 0.42	&   -6.56   &0.41 &  	-6.56	&0.41 &  13    \\ 
\ion{Ni}{ii}	&-6.06  	&	 0.56	&   -6.17   &0.58 &  	-6.17	&0.58 &  2     \\ 
\ion{Cu}{i}	&-8.29  	&	 0.31	&   -8.40   &0.32 &  	-8.40	&0.33 &  2     \\ 
\ion{Zn}{i}	&-8.16  	&		&   -8.27   &	  &  	-8.27	&     &  1     \\ 
\ion{Sr}{i}	&-6.94  	&	 0.19	&   $\ast$    &     &  	$\ast$	&     &  12    \\ 
\ion{Sr}{ii}	&-7.19  	&	 0.52	&   $\ast$    &     &  	$\ast$	&     &  2     \\ 
\ion{Y}{i}	&-8.48  	&	 0.08	&   -8.59   &0.08 &  	-8.58	&0.08 &  4     \\ 
\ion{Y}{ii}	&-8.80  	&	 0.22	&   -8.91   &0.23 &  	-8.90	&0.24 &  11    \\ 
\ion{Zr}{ii}	&-9.38  	&	 0.17	&   -9.51   &0.18 &  	-9.51	&0.18 &  5     \\ 
\ion{Ba}{ii}	&-9.93  	&		&   -10.05  &	  &  	-10.05  &     &  1     \\ 
\ion{La}{ii}	&-10.38 	&	 0.24	&   -10.54  &0.22 &  	-10.54  &0.25 &  3     \\ 
\ion{Ce}{ii}	&-9.81  	&	 0.27	&   -9.97   &0.27 &  	-9.97	&0.27 &  6     \\ 
\ion{Ce}{iii}	&-6.75  	&	 0.34	&   -6.88   &0.37 &  	-6.84	&0.37 &  5     \\ 
\ion{Pr}{ii}	&-9.36  	&		&   $\ast$    &	  &  	$\ast$	&     &  1     \\ 
\ion{Pr}{iii}	&-9.04  	&	 0.19	&   $\ast$    &     &  	$\ast$	&     &  4     \\ 
\ion{Nd}{ii}	&-10.27 	&	 0.36	&   $\ast$    &     &  	$\ast$	&     &  9     \\ 
\ion{Nd}{iii}	&-7.32  	&	 0.05	&   $\ast$    &     &  	$\ast$	&     &  2     \\ 
\ion{Sm}{ii}	&-9.39  	&	 0.24	&   -9.54   &0.24 &  	-9.54	&0.24 &  7     \\ 
\ion{Sm}{iii}	&-6.92  	&	 0.13	&   -6.94   &0.15 &  	-6.91	&0.16 &  4     \\ 
\ion{Eu}{ii}	&-9.91  	&	 0.17	&   -10.06  &0.17 &  	-10.04  &0.17 &  4     \\ 
\ion{Gd}{ii}	&-9.01  	&	 0.60	&   -9.16   &0.60 &  	-9.16	&0.61 &  11    \\ 
\ion{Gd}{iii}	&-7.65  	&		&   -7.87   &	  &  	-7.86	&     &  1     \\ 
\ion{Tb}{iii}	&-8.88  	&	 0.31	&   -9.08   &0.31 &  	-9.07	&0.31 &  7     \\ 
\ion{Dy}{ii}	&-9.96  	&	 0.37	&   -10.07  &0.36 &  	-10.07  &0.36 &  12    \\ 
\ion{Dy}{iii}	&-7.45  	&	 0.26	&   -7.63   &0.26 &  	-7.62	&0.26 &  4     \\ 
\ion{Ho}{ii}	&-10.19 	&		&   -10.26  &	  &  	-10.27  &     &  1     \\ 
\ion{Ho}{iii}	&-8.38  	&	 0.05	&   -8.58   &0.06 &  	-8.57	&0.06 &  3     \\ 
\ion{Er}{ii}	&-9.89  	&	 0.85	&   -10.02  &0.84 &  	-10.02  &0.83 &  2     \\ 
\ion{Er}{iii}	&-8.10  	&		&   -8.28   &	  &  	-8.28	&     &  1     \\ 
\ion{Tm}{ii}	&-10.84 	&	 0.22	&   -10.97  &0.20 &  	-10.97  &0.20 &  2     \\ 
\ion{Tm}{iii}	&-8.19  	&		&   -8.41   &	  &  	-8.40	&     &  1     \\ 
\ion{Yb}{ii}	&-9.76  	&	 0.46	&   -9.89   &0.47 &  	-9.89	&0.47 &  3     \\ 
\ion{Lu}{ii}	&-10.38 	&	 0.14	&   -10.53  &0.14 &  	-10.52  &0.14 &  2     \\ 
\hline
\end{tabular}
\end{scriptsize}
\end{table}
\end{center}

\begin{figure}
\includegraphics[angle=270, width=0.5\textwidth]{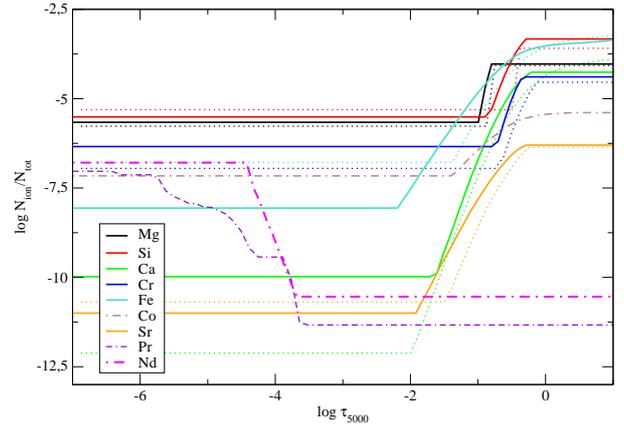}
\caption{Stratification profiles of the final model with$\teff$= 7550 K and $\log g$= 3.8 (solid lines) and initial homogeneous 
model (dotted lines). Dash-dotted lines indicate elements that were not included in the initial model but added at a later stage.}
\label{STRATFirstVSFinal}
\end{figure}

The final converged model contains stratification profiles of nine elements, including REEs. Only for Pr and Nd a chemical gradient with increased abundances in upper
atmospheric layers was necessary to fit the observed spectral line profiles. All other elements indicate opposite gradients with concentration in deeper layers. A
summary of the stratification profiles of all elements in the final model is shown in Figure \ref{STRATFirstVSFinal}.
Stratification profiles did not change after convergence was reached, compared to the initial homogeneous model. 

The largest differences for upper atmospheric abundances, which were found to be the most variable of the four parameters during the iterative process, 
were observed for Ca and Fe.

Detailed results of the \ddafit\ fitting routine, including formal errors for all four derived stratification parameters (upper or lower atmospheric abundance, position and
width of the abundance jump in the column mass scale used for calculations) are presented in Table~\ref{StratResults}.

\begin{figure*}
\includegraphics[width=0.5\textwidth]{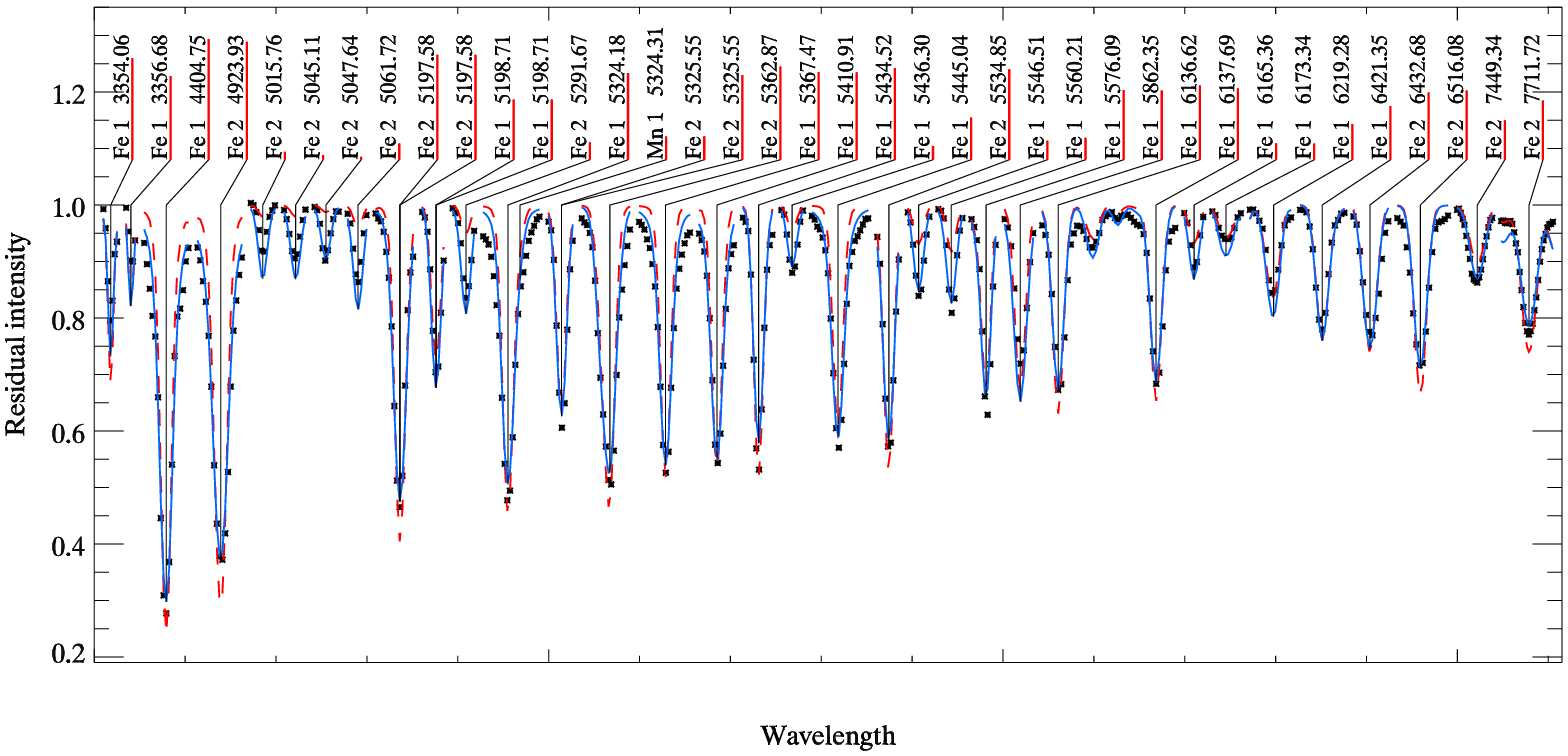}
\includegraphics[width=0.5\textwidth]{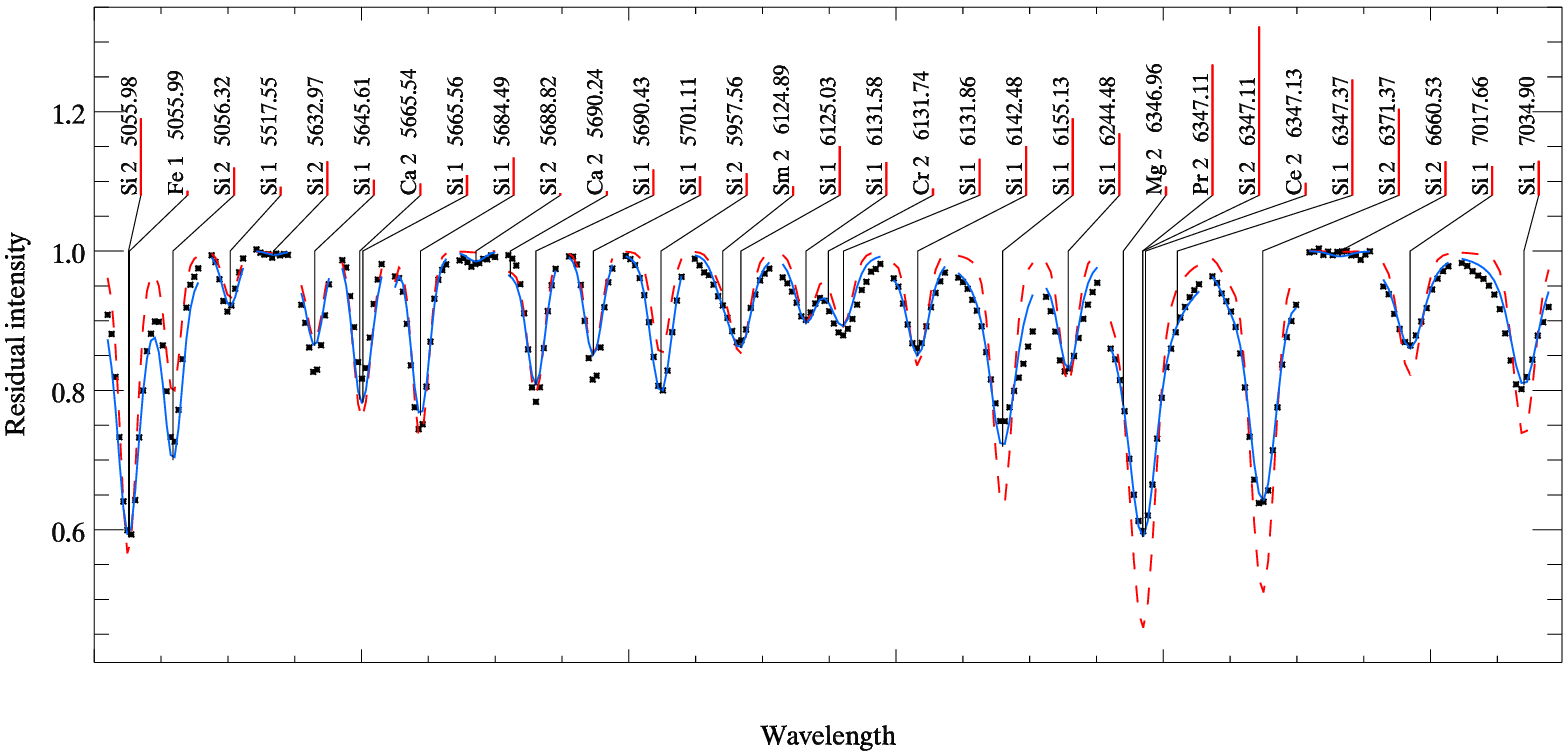}
\caption{Fit of observed (black) and synthetic lines with homogeneous (red) and stratified (blue) abundances. Fe is shown in the left column, Si in the right column. }
\label{ddafitLines}
\end{figure*}

Two examples of observed lines, synthetic profiles with the best fitting homogeneous abundance 
and stratified abundances are shown for Fe and Si in Fig.~\ref{ddafitLines}.
Additional figures for other elements are included in the online material. 
Deviation between observed and final synthetic line profiles were between 1.7\% and 3.4\% for all elements.

\begin{table}
\caption{Stratification parameters for 7 elements based on the final model. Upper
and lower atmospheric abundance values, abundance step position and width in column
mass scale and
deviation between observed and synthetic spectra are given. Formal errors of the \ddafit\ procedure are given. }
\label{StratResults}
\begin{tabular}{lccccc}
\hline\hline
       & log$N/N_{\rm up}$     & log$N/N_{\rm low}$    & log$M_{\rm step}$   & $\Delta$ log$M_{\rm step}$ & dev\\
       & $\pm$		       & $\pm$		       &       $\pm$		       &       $\pm$     &	  \\
\hline
Mg     &       -5.66	&	-4.03   &	  0.06  &	 0.04	&	2.8 \%  \\
       &        0.10	&	 0.02   &	  0.03  &	 0.54	&		 \\
Si     &       -5.51	&	-3.33   &	  0.17  &	 0.16   &	1.7 \%  \\
       &        0.09	&	 0.10   &	  0.01  &	 0.06   &		 \\
Ca     &       -9.98    &       -4.26	&	 -0.02  &	 0.56   &	3.2 \%  \\
       &        0.59    &        0.07	&	  0.01  &	 0.03   &		 \\
Cr     &       -6.30	&       -4.42	&	  0.21 	&	 0.02	&	2.4 \%  \\
       &        0.06	&        0.10	&	  0.01	&	 0.10	&		\\
Fe     &       -8.06    &       -3.16	&        -0.14  &        0.95   &       2.6 \%  \\
       &        0.78    &        0.28	&         0.04  &        0.19   &	       \\
Co     &       -7.16	&       -5.39	&	  0.06  &        0.46   &       3.4 \%  \\
       &        0.16	&        0.41	&	  0.05  &        0.36   &	       \\
Sr     &      -11.00    &	-6.30   &	 -0.11  &	 0.70   &       2.0 \%  \\
       &        0.73   &	 0.22   &	  0.04  &	 0.17   &		   \\
\hline
\end{tabular}
\end{table}

\subsubsection{Magnesium}

Magnesium stratification profiles did not change significantly between the first and final iterations 
and were not sensitive to small atmospheric structure changes in any region. 
Analysis of the contribution functions indicates that the line set is sensitive to a variety of atmospheric depths, 
as it includes two rather strong \ion{Mg}{ii} lines in the UV as well
as a number of strong infrared lines, in addition to strong lines in the optical spectral region. 
The final stratification profile shows an abundance jump of 1.7~dex which
is needed to reproduce observed line profiles.

\subsubsection{Silicon}

A variety of suitable Si lines were available for stratification analysis including five high excitation \ion{Si}{ii} lines useful 
for probing deeper atmoshperic 
layers, two strong \ion{Si}{ii} lines sensitive to upper atmospheric abundances, two infrared \ion{Si}{i} lines as well as a number 
of \ion{Si}{i} and \ion{Si}{ii} lines that exhibit broad line wings. 
Throughout the iterative fitting process, the Si distribution remained stable between different iterations. 
The abundance jump of 2~dex is located in the same optical depth as for the Cr profile.  

\subsubsection{Calcium}

Calcium shows the largest abundance stratification of all elements investigated in \aql\ with an abundance change of 5.7~dex 
between lower and upper atmospheric layers. The range
of the transition region of the step profile was found to be $\log\tau_{\rm5000}=[-0.2,-1.7]$. The upper atmospheric abundance was very 
sensitive to small changes in atmospheric structure
occuring during the iterative model calculations, before reaching a converged
solution. As many strong Ca lines with anomalously strong wings and shallow line
cores were included in
the analysis, the lower boundary of the abundance jump was constrained with better accuracy 
than the upper atmospheric abundance.

\subsubsection{Chromium}

Like Fe and Si, a variety of spectral lines was available for Cr analysis, including
lines with high excitation energies, 
and strong UV lines which were found to be very sensitive to high
atmospheric regions up to $\log\tau=-2$ according to the calculated contribution functions. 
The derived abundance profile remained very stable over all iterations and shows a small
abundance gradient of 2~dex occuring in the same depth region as the Si abundance step. 

\subsubsection{Iron}

Of all the investigated elements the highest number of suitable spectral lines for stratification calculations was found for Fe. 
Therefore the line list is expected to sample a large depth range of the stellar atmosphere. Two weak UV lines, two infrared lines, 
strong and weak lines of different ionization stages as well as high excitation
\ion{Fe}{ii} lines were included and a large final abundance gradient of 4.6~dex was derived. Inspection of the calculated 
contribution functions however revealed that only two strong lines
were sensitive to higher atmospheric layers around $\log\tau=2$ while most other lines were mainly 
found to be formed between $\log\tau=-0.3$ and $-1.5$ in a stratified atmosphere.
The upper atmospheric abundance was therefore very sensitive to small changes in atmospheric structure 
between different iteration steps. Nevertheless the lower atmospheric 
abundance, the position, and slope of the abundance gradient did not change significantly between first and final iterations. 

\subsubsection{Cobalt}

The stratification analysis of Co became possible with the availability of better atomic line data, allowing to take 
into account hyperfine structure in spectrum
synthesis. The coresponding data were taken from \citet{1996ApJS..107..811P} (\ion{Co}{i}) and \citet{2010MNRAS..401.1334B} 
(\ion{Co}{ii}). This is especially important as Co 
stratification is expected to be present (as indicated by the results of the abundance analysis) 
but not very strong. The final Co stratification was best represented by an abundance jump of 1.8~dex. 
This small abundance gradient (which was based on a short line list including strong UV lines and intermediate 
and weak optical lines) proved not to be very sensitive to small changes in model structure.

\subsubsection{Strontium}

Formation depth calculations predicted that the available set of Sr lines would be sensitive to a very narrow atmospheric region.
Upper atmospheric abundance values were therefore
found to be highly dependent on small changes in model structure and choice of input
parameters for the fitting routine,
throughout the iterative stratification fitting procedure. 
The final stratification profile of the converged model, however, did not
differ significantly from the initial one. A large abundance gradient of 4.7~dex was fit to the observed line profiles 
by our algorithm. However, as the upper atmospheric 
abundance is constrained by only a few points in the line profiles of the strongest lines, 
the real size of the abundance jump might be overestimated. 

\subsubsection{Rare earth elements}

\begin{figure}
\includegraphics[width=0.5\textwidth]{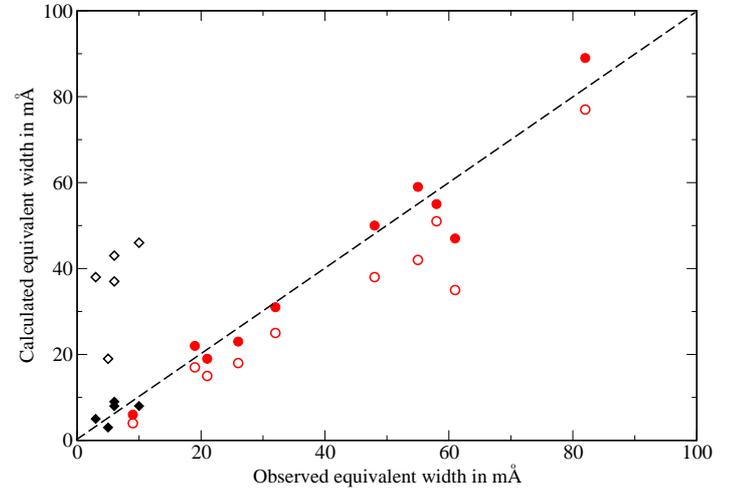}
\caption{Comparison between the observed and calculated equivalent widths of \ion{Nd}{ii} (diamonds) and \ion{Nd}{iii} (circles) lines. 
Calculations are made with the derived Nd stratification in LTE (open symbols) and NLTE (filled symbols) approximations. }
\label{REE}
\end{figure}

For {Pr} and {Nd} we performed a NLTE stratification analysis as described in \citet{2005A&A...441..309M,2009A&A...495..297M} 
using a trial-and-error method and the observed equivalent widths of the lines of the first and second ions, therefore the shape of the abundance profile may differ from a single-step function.
Possible magnetic intensification for strong \ion{Nd}{iii} lines was approximated
by pseudomicroturbulent velocity of 1.0~km/s. 
The observed and calculated equivalent widths of Nd lines are illustrated in Fig.~\ref{REE}.

Pr and Nd are the only elements which are enriched in upper atmospheric layers between $\log\tau=-3.6$ and $-6.0$ 
and therefore required inclusion of inverted step 
profiles, compared to the other elements. The observed differences between upper and lower atmospheric abundances 
were 3.7~dex for Nd and 4.2~dex for Pr. 


\subsection{Fundamental and atmospheric parameters}

\begin{figure*}[h!]
\includegraphics[width=\hsize]{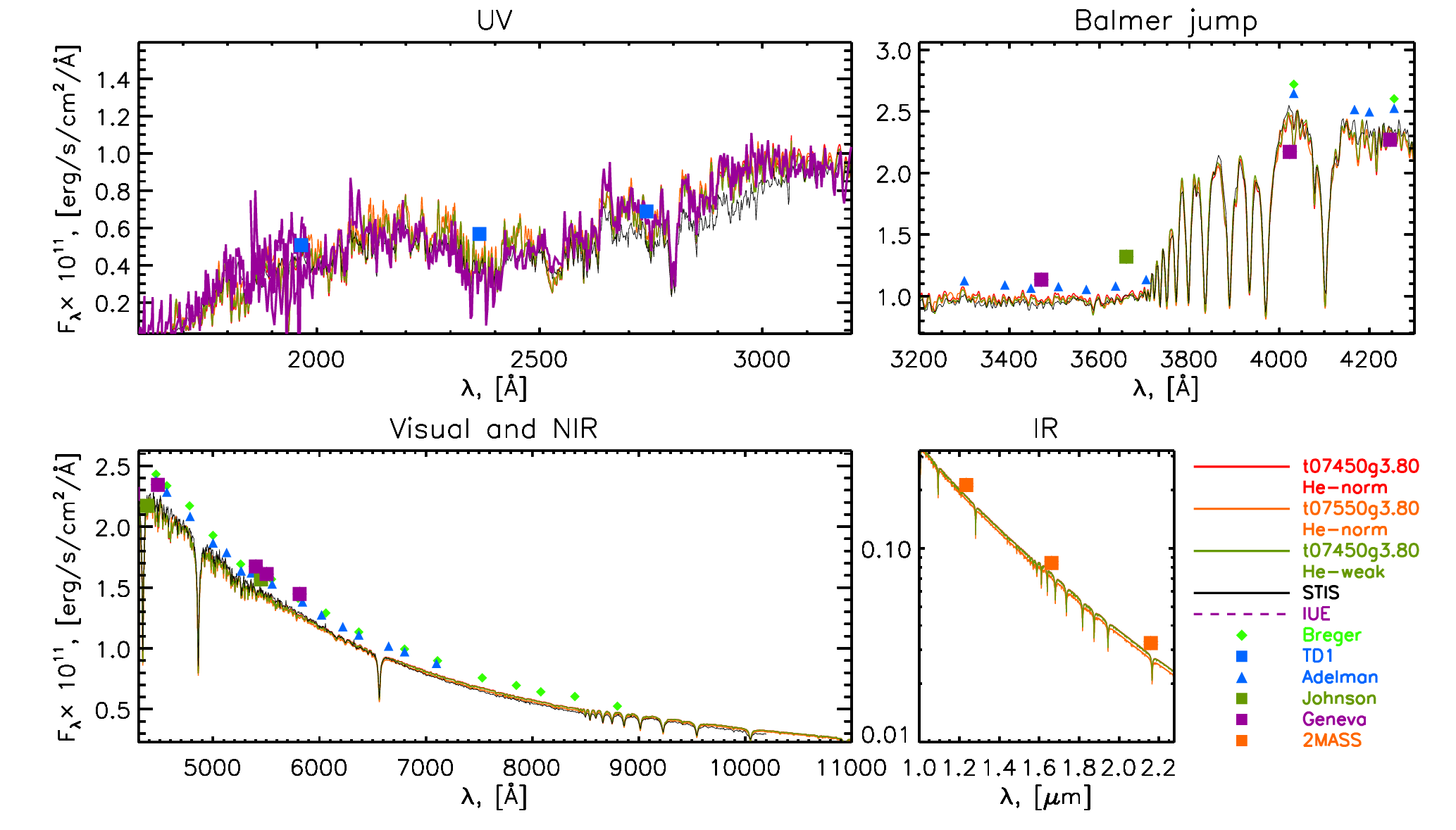}
\caption{Comparison between observed and predicted fluxes of \aql. The model fit is done to the data obtained with
STIS instrument. Please, see the online version of this figure for color-coding.}
\label{FluxFinal-1}
\end{figure*}

Our determination of atmospheric parameters is based on the model fit to the observed energy distributions
calibrated to absolute units (see Sect.~\ref{sec:isc}). We note a discrepancy between space- and ground-based observations
as can be seen from Fig.~\ref{FluxFinal-1}. In particular, data taken from the STIS archive are systematically lower than
that of, e.g., broad-band spectrophotometry of \citet{1989A&AS...81..221A} and \citet{1976ApJS...32....7B} as well as fluxes 
computed from Johnson and Geneva
photometric systems. In the attempt to fit these datasets one additional assumption about atmospheric chemistry was made.
As follows from the computations of particle diffusion, helium always sinks 
in subphotosheric layers of A-F stars \citep{1979ApJ...234..206M}. The same He depletion was found later also
in the optically thin layers \citep{2004IAUS..224..193L} suggesting that the atmospheres of CP stars of spectral
types A-F may well be He-deficient. Having no spectroscopic clues about the true He stratification, we thus considered
both He-normal (i.e. solar) and He-weak atmospheres with adopted He abundance of $\log(He/H)=-4$ 
(decreasing He abundance below this value has only marginal or no effect at all).

In addition, using observed spectral energy distributions and stellar parallaxes allowed us to derive
the radius of the star. The latter is obtained by minimizing the deviation between observed and predicted fluxes for a given
$\teff$ and $\logg$. The parallax $\pi=12.76\pm0.29$~mas was taken from \citet{leeuwen}.

Making use of STIS data resulted in $\teff=7450$~K, $\logg=3.8$ both for He-normal and He-weak models. 
Formally, the He-weak model with $\logg=3.6$ provided a better fit with slightly lower $\chi^2$ compared to the model with $\logg=3.8$,
but the computed mass of the star is then found to be $0.93\Msun$ which is too small for an A-type star.
Furthermore, the Paschen continuum, which is a good indicator of the atmospheric temperature, 
is clearly requiring a hotter temperature, which we find to be $\teff=7550$~K.
The examples of the model predictions are shown on Fig.~\ref{FluxFinal-1}.
The respective fundamental stellar parameters are:
$R=2.53\pm0.06\Rsun$, $M=1.47\pm0.07\Msun$, $L=17.78\pm0.85\Lsun$ for $\teff=7450$~K model
and $R=2.46\pm0.06\Rsun$, $M=1.39\pm0.07\Msun$, $L=17.73\pm0.88\Lsun$ for the $\teff=7550$~K model respectively.
Note that the error-bars results from the parallax accuracy only.

\begin{figure*}
\includegraphics[width=\hsize]{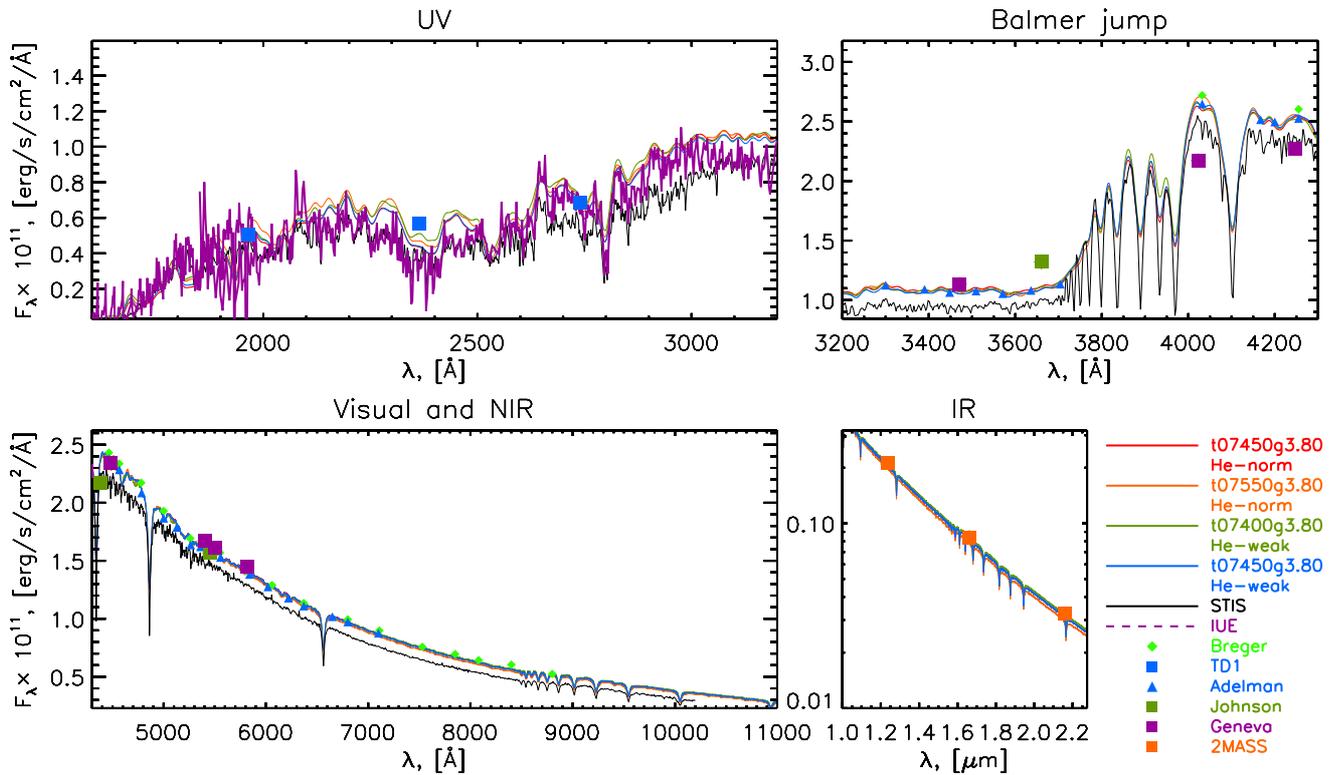}
\caption{Same as on Fig.~\ref{FluxFinal-1} but with the model fit to the ground-based spectrophotometric data
of  \citet{1989A&AS...81..221A} and \citet{1976ApJS...32....7B}.}
\label{FluxFinal-2}
\end{figure*}

As a next step we fitted the data obtained from ground-based observations. For this purpose we used spectrophotometric 
observations by \citet{1989A&AS...81..221A} and extended them by NIR points ($\lambda>7000$\AA) from \citet{1976ApJS...32....7B}.
We did so because observations of Adelman cover the region of the Balmer jump and
thus provide a homogeneous (in the sense of calibration) 
set of points in this important spectral region.
In this particular case we find that He-normal models with
$\teff=7450-7550$~K, $\logg=3.8$ and He-weak models with
$\teff=7400-7450$~K, $\logg=3.8$ provide the best fit. The comparison of the
observed and predicted fluxes is
shown in Fig.~\ref{FluxFinal-2}.
The respective fundamental stellar parameters for the He-normal and He-weak
settings are given in table \ref{Heparam}.

\begin{table}
\caption{Fundamental parameters for He-normal and He-weak models with $\teff$ corresponding to the best fits to observed fluxes.}
\label{Heparam}
\begin{center}
\begin{tabular}{lcc}
\hline\hline
	&\multicolumn{2}{c}{He-normal}\\
$\teff$	&7450 K&7550 K\\
\hline
R &	$2.67\pm0.06\Rsun$	&	$2.59\pm0.06\Rsun$	 \\
M &	$1.64\pm0.08\Msun$	&	$1.55\pm0.07\Msun$	 \\
L &	$19.81\pm0.90\Lsun$	&	$19.66\pm0.90\Lsun$	 \\
\hline
	&\multicolumn{2}{c}{He-weak}\\
$\teff$	&7400 K&7450 K\\
\hline
R	&	$2.72\pm0.06\Rsun$	&	$2.67\pm0.06\Rsun$	\\
M	&	$1.70\pm0.08\Msun$	&	$1.64\pm0.08\Msun$	\\
L	&	$20.01\pm0.90\Lsun$	&	$19.81\pm0.90\Lsun$	\\
\hline
\end{tabular}
\end{center}
\end{table}





\begin{figure*}
\includegraphics[width=\hsize]{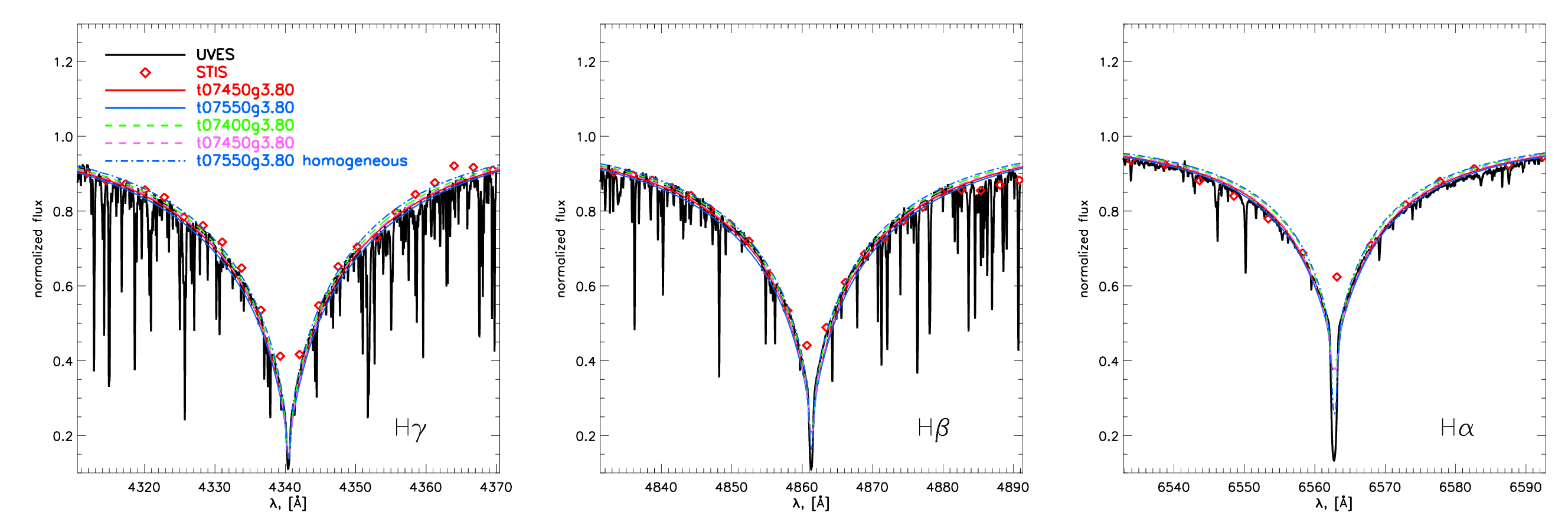}
\caption{Observed Hydrogen Balmer lines compared to synthetic spectra calculated with different sets of fundamental parameters used during the iterative process.
The dashed lines correspond to models computed with depleted He content.}
\label{HLines_Denis}
\end{figure*}

Figure~\ref{HLines_Denis} illustrates a fit to the observed hydrogen Balmer line profiles for several models. It is seen
that generally a model with $\teff=7450$~K provides an optimal fit to all three Balmer lines. On the other hand
$\hbeta$ and $\hgamma$ are best fitted with a slightly cooler $\teff=7400$~K He-weak model.

Taking into account a substantial scatter in the observed energy distributions, the derived atmospheric
parameters of the star are $\teff=7500\pm50$~K, $\logg=3.8\pm0.1$. There are two major spectral regions that play a crucial
role in the determination of atmospheric parameters: the Balmer jump (which controls best the value of surface gravity) and the
Paschen continuum (which controls best the effective temperature). Therefore, it is important to use observed fluxes
from the same data source that cover these two regions. Different
calibration schemes applied by different observers could result in a systematic
scatter in the derived model
parameters. For instance, fluxes obtained by Breger lay slightly above the fluxes by
Adelman in the region of the Balmer jump
(see, e.g., top right panel of Fig.~\ref{FluxFinal-1}). Using them entirely in model fitting together with IUE data
or UV points of Adelman results in a hotter best-fit temperature of
$\teff=7550-7600$~K, compared to the one found by fitting
data of Adelman extended with NIR points of Breger. Although this is not a large discrepancy, still one should be
aware of similar systematics when analysing other stars for which no homogeneous datasets are available.

The difference between space- and ground-based observations results in radius uncertainity of
$\Delta R=0.14\Rsun$ (assuming the same $\teff$ for reference models). This is a rather small discrepancy, which is
of the same order as the errors provided by modern interferometry, as found by, e.g., \citet{2011A&A...526A..89P}
for another bright Ap star $\gamma$~Equ. Still, with improving interferometric accuracy, such a scatter between
different observed spectrophotometric datasets could become important.

\begin{figure}
\includegraphics[angle=270, width=\hsize]{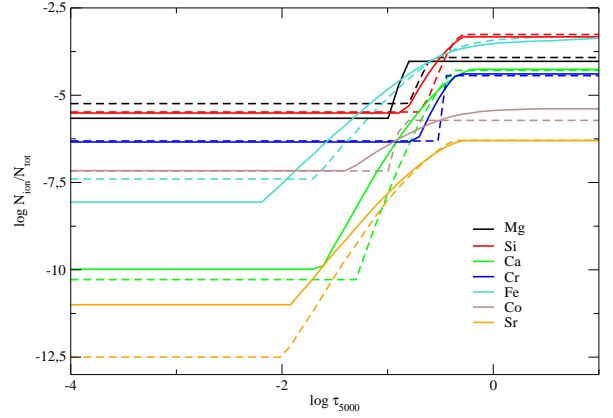}
\caption{Comparison of stratification profiles of 7 elements computed with He-normal (solid lines) or He-weak (dashed lines) atmosphere models.}
\label{HeStrat}
\end{figure}

To evaluate the influence of a change in model structure with modified He abundance on the derived stratification, 
two sets of profiles were computed using the
He-weak and He-normal models with $\teff=7550$~K and $\log g=3.8$. Results of this comparison are shown in Fig.~\ref{HeStrat}.
A small systematic shift of abundance jumps towards
lower atmospheric layers is observed when a He-weak composition is assumed. The quality of the fit of synthetic line profiles
to observations was not found to be different between the two models.

\begin{figure}
\includegraphics[width=\hsize]{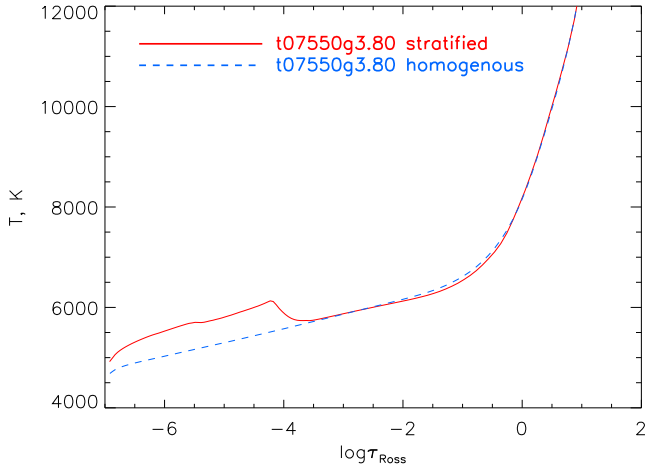}
\caption{Temperature structure of models computed with homogeneous and stratified abundance patterns.}
\label{fig:t}
\end{figure}

\section{Discussion and conclusions}
\label{sec:concl}
We have constructed a model atmosphere for \aql\ using a self-consistent iterative approach to derive atmospheric 
abundances and vertical chemical gradients by fitting
models to the observed spectral lines and wide-range energy distributions.

By analyzing line profiles in a single spectrum we are able to
derive somewhat mean vertical abundance gradients, i.e. gradients
averaged over the visible surface of a star. It is to note that
modern Doppler Imaging techniques (that rely on the rotationally
modulated variability of line profiles) provide us with
two-dimensional images of stellar surfaces and reveal non-uniform
horizontal distributions of chemical elements and magnetic fields in
atmospheres of CP stars. It is therefore natural to think of
horizontal inhomogeneities as a result of variable vertical gradients, which 
depend on the local magnetic field geometry, hydrodynamic flows,
and possibly other processes (that we don't yet fully understand), the
resulting force that pushes ions upwards or makes them diffuse
downwards in the stellar atmosphere becomes variable over the stellar
surface. That means that the positions of abundance jumps as well as
their amplitudes are then the functions of local surface coordinates
too. Only a self-consistent 3D mapping of stellar
atmospheres has the potential of providing vertical and horizontal
abundance gradients simultaneously from the same observed data sets.
Modern theoretical models of atomic diffusion indeed predict a rather
strong sensitivity of element  stratification profiles to the geometry
of the surface magnetic field \citep[see, for
example][]{2009A&amp;A...495..937L,2010A&amp;A...516A..53A,2012MNRAS.425.2715S}.

However, the sharpness of spectral lines and lack of prominent line profile and magnetic field variations
imply a very long rotational period for 10 Aql \citep{Ryabchikova2005,2008MNRAS.389..903S}. In
this situation we are not able to use Doppler imaging to constrain surface distributions of
chemical elements. Very weak long-term line profile variability detected by \citep{2008MNRAS.389..903S}
suggests that horizontal abundance gradients in this star are not large or that the star is viewed
from nearly the same aspect angle during its rotation cycle. We have estimated 
that this weak variability would result in a mean abundance difference of $\le$0.05 dex. Therefore, we do not expect
stratification analysis to be significantly affected by horizontal inhomogeneities.

Our approach of using step-like assumption on derived abundance
gradients has obvious limitations. First of all it is limited to a
very narrow atmospheric region where a given set of atomic lines used
in stratification analysis is formed. It naturally cannot predict
abundances in optically very thin and thick layers which we simply do
not see with our instruments. Secondly, it is impossible to restore any
kind of complex vertical distribution whose shape may strongly
deviate from a step-like assumption \citep[see, for
example][]{2009A&amp;A...495..937L,2010A&amp;A...516A..53A,2012MNRAS.425.2715S}.
However, what is important for us
is to see a general trend of element distributions as a function of
depths, i.e. whether a given element is being brought upwards or
downwards in the line forming region of a stellar atmosphere and how
strong the resulting abundance gradient is. In spite of its
limitations this approach is suitable of tracking any kind of
systematic as applied to different stars with, e.g., different
temperatures and surface magnetic fields.

The detailed analysis of the high resolution spectra ranging from UV to infrared regions indicated that robustness of the inferred
 vertical stratification profiles is very much dependent on the sensitivity of the selected line set 
to a wide range of atmospheric depths. As seen, e.g. in 
our analysis of Cr stratification, inclusion of
high- and low-excitation lines helps
to constrain parameters describing a chemical gradient best. However, upper atmospheric abundances are sometimes 
still only constrained by a few lines which are sensitive to 
high atmospheric layers. Therefore the value of the upper atmospheric abundance might vary between 
an initial homogeneous model atmosphere and a stratified model found after the
iterative calculations have converged. 

The mean abundances for non-stratified elements changed by about 0.1~dex between 
the homogogeneous starting model and all other stratified models. 
Abundance variations between different stratified models were negligible.

The most striking change between our homogeneous starting model and the final stratified model atmosphere 
is seen in a systematic shift of the abundance jumps towards higher
atmospheric layers in the stratified case. 

Between the first and final models, derived stratification profiles were used to calculate synthetic flux distributions 
to be compared with a large number of observations. 
This comparison was used for fine tuning of fundamental parameters of \aql. 
We find that stratification itself has little effect on the synthetic energy distribution of the star compared
to models computed with individual and homogeneous abundances. The overall fit and derived radii were almost identical.
Therefore individual abundances appear to be the most important ingredient in the SED modelling, at least
in the roAp temperature region. On the other hand, we do find that the observed hydrogen lines could not be reproduced 
with chemically homogeneous model atmospheres, as illustrated on Fig.~\ref{HLines_Denis} where we also show two
He-normal models with $\teff=7550$~K, $\logg=3.8$ computed with stratified and homogeneous abundances.
Therefore, the inclusion of vertical
element stratification has to be preferred when trying to establish an atmosphere model for a single Ap star 
in order to reproduce spectroscopic and SED observation simultaneously.

Similar to the case of HD~24712 \citep{2009A&A...499..879S} we find a strong influence of REE opacity in the
surface atmospheric structure of stratified models. In particular, strong abundance gradients of Pr and Nd 
shown in Fig.~\ref{STRATFirstVSFinal} lead to a heating of plasma and appearance of the characteristic temperature jump. 
This is illustrated on Fig.~\ref{fig:t} for He-normal final models. Interesting to note that such a temperature
jump was first empirically predicted by \citet{2002ApJ...578L..75K} in order to fit narrow region of Balmer line
profiles between wings and core (the so-called core-wing anomaly). Unfortunately,
the temperature jump predicted by our models
is located way too high in the atmosphere compared to what was suggested by \citet{2002ApJ...578L..75K} (see their Fig.~3).
We hope that a self consistent NLTE modelling of REE stratification will help to improve existing models and finally
match the predicted position of the temperature jump. Last but not least, the
incorporation of an atmospheric model
with an inverse temperature gradient from \citet{2009A&A...499..879S} allowed to improve pulsation models of HD~24712
as recently computed by \citet{2010MNRAS.403.1729S}. All this suggests that the temperature structure of Ap stars
is indeed very much different from the ``canonical''  $T-\tau$ relation and that the derived temperature jumps are real.

One of the important results of the present study is the decrease of the $\logg$ value from $4.0$ to $3.8$
once the fitting of SED is introduced in the atmospheric analysis. This is caused by the incorporation of
realistic chemistry in model atmosphere computations as well as the use of homogeneous spectral energy
distributions (i.e. STIS data in this particular case) which cover the whole Balmer jump region.
A similar decrease of surface gravity was also required in two previous investigations of Ap stars
$\alpha$~Cir \citep{2009A&A...499..851K} and HD~24712 \citep{2009A&A...499..879S}.
In the latter case, recent pulsation modelling by \citet{2010MNRAS.403.1729S} also suggested that
$\logg$ of HD~24712 should be lower than what we usually find from photometric calibrations of Ap stars.
From our analysis we conclude  that a decrease of $\logg$ is consistent with all
observations if the appropriate abundance pattern is included in model calculations.
On the other hand, decreasing surface gravity does not dramatically change stratification profiles 
(see our first and final profiles in Fig.~\ref{STRATFirstVSFinal}).

One of the open questions that still needs to be answered is the true helium content in the atmospheres of Ap stars.
Using available observational material it is impossible to distinguish between He-normal and He-weak models 
with high accuracy. Surface temperatures of roAp stars are too low to see He lines.
The helium concentration formally 
influences the quality of the fit to energy distributions, but once plotted against low and moderate resolution 
data the difference is difficult to see, except for the Balmer jump region. 
In addition, hydrogen line profiles are also only marginally affected by He depletion.
Pulsation calculations may therefore assume He depletion and still be compatible with all available observations 
(spectroscopy, spectrophotometry). Again, improved self-consistent diffusion models are probably
the right way to go, in the search for a definite answer.

\begin{acknowledgements}
Part of this work was supported by the Austrian Fonds zur F\"orderung wissenschaftlicher Forschung, project P17890 and P22691.
DS is granted by Deutsche Forschungsgemeinschaft (DFG) Research Grant RE1664/7-1.
OK is a Royal Swedish Academy of Sciences Research Fellow, supported by the grants from Knut and Alice Wallenberg Foundation and Swedish Research Council.

We also acknowledge the use of electronic databases (VALD, SIMBAD, NASA's ADS) and 
cluster facilities at Vienna Institute for Astronomy and Georg August University G\"ottingen.
\end{acknowledgements}


\longtab{3}{
\begin{longtable}{l|l|l|l|l|l}
\caption{Atomic parameters for all lines used in stratification analysis of 10~Aql.
Abbreviations of references correspond to different internal line lists of VALD and individual line lists with observed parameters. 
In case broadening constants were absent in VALD, approximations were used. 
\label{atomicparsALL}
}\\
\hline
\hline
 Ion &	$\lambda$ [\AA]   &$\log gf$            & $E_{\rm low}$  &$\gamma_{\rm Stark}$  &Ref.      \\	 
\hline
\endfirsthead
\caption{continued.}\\
\hline 
\hline 
 Ion &	$\lambda$ [\AA]   &$\log gf$            & $E_{\rm low}$  &$\gamma_{\rm Stark}$  &Ref.      \\
\hline
\endhead
\caption{continued.}\\
\hline
\hline
 Ion &  $\lambda$ [\AA]   &$\log gf$            & $E_{\rm low}$  &$\gamma_{\rm Stark}$  &Ref.      \\
\hline
\endhead
\endfoot
\ion{Mg}{i}  &	  4702.99 & -0.420   & 4.346 &-4.460       & Bu      \\ 
\ion{Mg}{i}  &	  4730.02 & -2.409   & 4.346 &approx       & JK      \\ 
\ion{Mg}{i}  &	  5172.68 & -0.380   & 2.712 &-5.470       & Wi      \\ 
\ion{Mg}{i}  &	  5183.60 & -0.160   & 2.717 &-5.470       & Wi      \\ 
\ion{Mg}{i}  &	  5528.40 & -0.400   & 4.346 &-4.460       & Bu      \\ 
\ion{Mg}{i}  &	  5711.08 & -1.833   & 4.346 &approx       & Ku      \\ 
\ion{Mg}{i}  &	  8806.75 & -0.200   & 4.346 &approx       & $\odot$   \\ 
\ion{Mg}{i}  &   8923.56 & -1.650   & 5.394 &approx       & LL      \\ 
\ion{Mg}{ii} &   3104.71 & -0.030   & 8.864 &-3.970       & Ku      \\ 
\ion{Mg}{ii} &   3104.72 & -1.330   & 8.864 &-3.970       & Ku      \\ 
\ion{Mg}{ii} &   3104.80 & -0.190   & 8.864 &-3.970       & Ku      \\ 
\ion{Mg}{ii} &   4390.51 & -1.700   & 9.999 &-4.070       & Wi      \\ 
\ion{Mg}{ii} &   4390.57 & -0.530   & 9.999 &-4.070       & Wi      \\ 
\ion{Mg}{ii} &   4427.99 & -1.200   & 9.996 &-4.400       & Wi      \\ 
\ion{Mg}{ii} &   4433.98 & -0.900   & 9.999 &-4.400       & Wi      \\ 
\ion{Mg}{ii} &   4481.12 &  0.740   & 8.864 &-4.700       & Ku      \\ 
\ion{Mg}{ii} &   4481.15 & -0.560   & 8.864 &-4.700       & Ku      \\ 
\ion{Mg}{ii} &   4481.32 &  0.590   & 8.864 &-4.700       & Ku      \\ 
\ion{Mg}{ii} &   7877.05 &  0.390   & 9.996 &-4.540       & Wi      \\ 
\ion{Mg}{ii} &   8213.98 & -0.950   & 9.999 &-4.770       & $\odot$   \\ 
\hline
\ion{Si}{i}  &   5517.55 & -2.384   &	5.080 &8.380  	     & BR      \\ 
\ion{Si}{i}  &   5645.03 & -2.400   &	5.614 &approx 	     & NL      \\ 
\ion{Si}{i}  &   5665.56 & -2.040   &	4.920 &8.290  	     & LL      \\ 
\ion{Si}{i}  &   5684.49 & -1.650   &	4.950 &8.250  	     & NL      \\ 
\ion{Si}{i}  &   5690.43 & -1.870   &	4.930 &8.300  	     & NL      \\ 
\ion{Si}{i}  &   5701.10 & -2.050   &	4.930 &8.310  	     & NL      \\ 
\ion{Si}{i}  &   6125.02 & -1.513   &	5.610 &approx 	     & sH      \\ 
\ion{Si}{i}  &   6131.57 & -1.705   &	5.610 &approx 	     & sH      \\ 
\ion{Si}{i}  &   6131.85 & -1.685   &	5.610 &approx 	     & sH      \\ 
\ion{Si}{i}  &   6142.48 & -1.420   &	5.620 &approx 	     & BR      \\ 
\ion{Si}{i}  &   6155.13 & -0.770   &	5.619 &approx 	     & BR      \\ 
\ion{Si}{i}  &   6244.47 & -1.363   &	5.610 &approx 	     & sH      \\ 
\ion{Si}{i}  &   7017.66 & -1.070   &	5.870 &approx 	     & sH      \\ 
\ion{Si}{i}  &   7034.90 & -0.780   &	5.871 &8.140  	     & BR      \\ 
\ion{Si}{ii} &   5055.98 &  0.440   &10.074 &9.040  	     & AJPP    \\  
\ion{Si}{ii} &   5056.31 & -0.359   &10.074 &9.030  	     & NL      \\  
\ion{Si}{ii} &   5632.96 & -0.820   &14.186 &approx 	     & Wil     \\  
\ion{Si}{ii} &   5688.81 &  0.000   &14.186 &approx 	     & Wil     \\  
\ion{Si}{ii} &   5957.55 & -0.350   &10.067 &8.820  	     & AJPP    \\ 
\ion{Si}{ii} &   6347.10 &  0.290   &	8.121 &9.090  	     & Wil     \\  
\ion{Si}{ii} &   6371.37 & -0.080   &	8.121 &9.080  	     & NIST    \\ 
\ion{Si}{ii} &   6660.53 &  0.230   &14.504 &approx 	     & Wil     \\ 
\hline  
\ion{Ca}{i}  &   4226.72 &  0.244   &	0.000 &-6.030 	     & SG      \\ 
\ion{Ca}{i}  &   5857.45 &  0.240   &	2.933 &-5.420 	     & S        \\ 
\ion{Ca}{i}  &   5867.56 & -1.570   &	2.933 &-4.705 	     & S        \\ 
\ion{Ca}{i}  &   6162.17 & -0.090   &	1.899 &-5.320 	     & SN      \\ 
\ion{Ca}{i}  &   6163.75 & -1.286   &	2.521 &-4.998 	     & SR      \\ 
\ion{Ca}{i}  &   6169.04 & -0.797   &	2.523 &-4.997 	     & SR      \\ 
\ion{Ca}{i}  &   6169.56 & -0.478   &	2.526 &-4.994 	     & SR      \\ 
\ion{Ca}{i}  &   6455.59 & -1.340   &	2.523 &-6.072 	     & S        \\ 
\ion{Ca}{i}  &   6471.66 & -0.686   &	2.526 &-6.072 	     & SR      \\ 
\ion{Ca}{ii} &   5021.13 & -1.207   &	7.515 &-4.612 	     & TB      \\ 
\ion{Ca}{ii} &   6456.87 &  0.410   &	8.438 &-3.711       & TB      \\ 
\ion{Ca}{ii} &   8248.79 &  0.556   &	7.515 &-4.600       & TB      \\ 
\hline 
\ion{Cr}{i}  &   3355.73 & -1.107   &	3.847 & -5.118      & GF	\\ 
\ion{Cr}{i}  &   5065.91 & -1.340   &	2.708 & -6.209      & GF	\\ 
\ion{Cr}{i}  &   5110.74 & -1.289   &	2.709 & -6.210      & GF	\\ 
\ion{Cr}{i}  &   5204.50 & -0.208   &	0.941 & -6.154      & MFW	\\ 
\ion{Cr}{i}  &   5247.56 & -1.640   &	0.961 & -6.120      & MFW	\\ 
\ion{Cr}{i}  &   5296.69 & -1.400   &	0.983 & -6.120      & MFW	\\ 
\ion{Cr}{i}  &   5345.80 & -0.980   &	1.004 & -6.117      & MFW	\\ 
\ion{Cr}{i}  &   5368.53 & -1.033   &	3.847 & -6.104      & GF	\\ 
\ion{Cr}{i}  &   6762.41 & -0.555   &	5.280 & -5.246      & GF	\\ 
\ion{Cr}{i}  &   6762.42 & -0.092   &	5.282 & -5.259      & GF	\\ 
\ion{Cr}{ii} &   3308.13 & -1.094   &	4.920 & -6.627      & RU	\\
\ion{Cr}{ii} &   3311.93 & -0.510   &	4.156 & -6.648      & RU	\\
\ion{Cr}{ii} &   3338.80 &  0.053   &	6.792 & -5.671      & RU	\\
\ion{Cr}{ii} &   3355.65 & -0.809   &	6.747 & -5.670      & RU	\\
\ion{Cr}{ii} &   3378.33 & -1.045   &	3.104 & -6.716      & RU	\\
\ion{Cr}{ii} &   5024.52 & -1.733   &	6.285 & -6.652      & RU	\\
\ion{Cr}{ii} &   5085.65 & -2.215   &	5.871 & -6.643      & RU	\\
\ion{Cr}{ii} &   5210.82 & -3.259   &	4.042 & -6.720      & RU	\\
\ion{Cr}{ii} &   5210.86 & -2.941   &	3.758 & -6.584      & RU	\\
\ion{Cr}{ii} &   5232.49 & -2.360   &	4.071 & -6.639      & RU	\\
\ion{Cr}{ii} &   5379.79 & -2.081   &	7.331 & -6.558      & RU	\\
\ion{Cr}{ii} &   5678.39 & -1.496   &	6.484 & -6.627      & RU	\\
\ion{Cr}{ii} &   6418.90 & -1.915   &	6.686 & -6.603      & RU	\\
\hline 
\ion{Fe}{i}  &   3354.05 & -1.169   &	2.858 & -6.091      & V2	\\
\ion{Fe}{i}  &   3356.68 & -1.554   &	3.047 & -5.082      & GF	\\
\ion{Fe}{i}  &   4404.75 & -0.142   &	1.557 & -6.204      & V2	\\
\ion{Fe}{i}  &   5198.71 & -2.135   &	2.223 & -6.185      & V2	\\
\ion{Fe}{i}  &   5324.17 & -0.103   &	3.211 & -5.496      & V2	\\
\ion{Fe}{i}  &   5367.46 &  0.443   &	4.415 & -5.133      & V2	\\
\ion{Fe}{i}  &   5410.90 &  0.398   &	4.473 & -5.060      & V2	\\
\ion{Fe}{i}  &   5434.52 & -2.122   &	1.011 & -6.303      & V2	\\
\ion{Fe}{i}  &   5436.29 & -1.540   &	4.386 & -4.997      & V2	\\
\ion{Fe}{i}  &   5445.04 & -0.020   &	4.386 & -4.582      & V2	\\
\ion{Fe}{i}  &   5546.50 & -1.310   &	4.371 & -6.109      & V2	\\
\ion{Fe}{i}  &   5560.21 & -1.190   &	4.434 & -4.323      & V2	\\
\ion{Fe}{i}  &   5576.08 & -1.000   &	3.430 & -5.491      & V2	\\
\ion{Fe}{i}  &   5862.35 & -0.058   &	4.549 & -4.582      & GF	\\
\ion{Fe}{i}  &   6136.61 & -1.400   &	2.453 & -6.327      & V2	\\
\ion{Fe}{i}  &   6137.69 & -1.403   &	2.588 & -6.112      & V2	\\
\ion{Fe}{i}  &   6165.36 & -1.474   &	4.143 & -6.156      & V2	\\
\ion{Fe}{i}  &   6173.33 & -2.880   &	2.223 & -6.194      & V2	\\
\ion{Fe}{i}  &   6219.28 & -2.433   &	2.198 & -6.202      & V2	\\
\ion{Fe}{i}  &   6421.35 & -2.027   &	2.279 & -6.310      & V2	\\
\ion{Fe}{ii} &   4923.92 & -1.504   &	2.891 & -6.583      & RU	\\
\ion{Fe}{ii} &   5015.75 & -0.028   &10.348 & -5.287      & RU	\\
\ion{Fe}{ii} &   5045.11 & -0.002   &10.308 & -4.984      & RU	\\
\ion{Fe}{ii} &   5047.64 & -0.235   &10.308 & -4.976      & RU	\\
\ion{Fe}{ii} &   5061.71 &  0.284   &10.308 & -5.189      & RU	\\
\ion{Fe}{ii} &   5197.57 & -2.348   &	3.230 & -6.599      & RU	\\
\ion{Fe}{ii} &   5291.66 &  0.544   &10.480 & -5.468      & RU	\\
\ion{Fe}{ii} &   5325.55 & -3.324   &	3.221 & -6.603      & RU	\\
\ion{Fe}{ii} &   5362.86 & -2.616   &	3.199 & -6.666      & RU	\\
\ion{Fe}{ii} &   5534.84 & -2.865   &	3.245 & -6.601      & RU	\\
\ion{Fe}{ii} &   6432.68 & -3.687   &	2.891 & -6.687      & RU	\\
\ion{Fe}{ii} &   6516.08 & -3.432   &	2.891 & -6.686      & RU	\\
\ion{Fe}{ii} &   7449.33 & -3.488   &	3.889 & -6.668      & RU	\\
\ion{Fe}{ii} &   7711.72 & -2.683   &	3.903 & -6.666      & RU	\\    
\hline 
\ion{Co}{i}  &   3502.27	&  0.070	  & 0.432 & -6.300 	    & FMW	 \\
\ion{Co}{i}  &   3502.61	& -1.240	  & 0.174 & -6.374 	    & FMW	 \\
\ion{Co}{i}  &   4813.46	&  0.050	  & 3.216 & -5.627 	    & GF	 \\
\ion{Co}{i}  &   5342.69	&  0.690	  & 4.021 & -4.928 	    & GF	 \\
\ion{Co}{i}  &   5347.49	& -0.160	  & 4.149 & -4.952 	    & GF	 \\
\ion{Co}{i}  &   6082.42	& -0.520	  & 3.514 & -5.474 	    & GF	 \\
\ion{Co}{ii} &   3415.77	& -1.740	  & 2.203 & -6.675 	    & SLW	 \\
\ion{Co}{ii} &   3501.71	& -0.970	  & 2.203 & -6.673 	    & T83av	 \\
\hline  
\ion{Sr}{i}  &   4607.32 & -0.570   & 0.000 &		approx      & Bh	\\
\ion{Sr}{i}  &   4722.27 & -0.220   & 1.798 &		approx      & GC	\\
\ion{Sr}{i}  &   4741.91 & -0.410   & 1.775 &		approx      & GC	\\
\ion{Sr}{i}  &   4784.31 & -0.510   & 1.798 &		approx      & Bh	\\
\ion{Sr}{i}  &   4811.87 &  0.190   & 1.847 &		approx      & GC	\\
\ion{Sr}{i}  &   5222.19 & -0.380   & 2.251 &		approx      & GC	\\
\ion{Sr}{i}  &   5229.26 & -0.330   & 2.259 &		approx      & GC	\\
\ion{Sr}{i}  &   5450.83 & -0.340   & 2.259 &		approx      & GC	\\
\ion{Sr}{i}  &   5486.13 & -0.460   & 2.251 &		approx      & GC	\\
\ion{Sr}{i}  &   5504.17 &  0.090   & 2.259 &		approx      & GC	\\
\ion{Sr}{i}  &   5521.76 & -0.060   & 2.251 &		approx      & GC	\\
\ion{Sr}{i}  &	  5540.04 & -0.410   & 2.259 &		approx      & GC	\\
\ion{Sr}{i}  &	  6408.45 &  0.510   & 2.271 &		approx      & GC	\\
\ion{Sr}{i}  &	  6503.99 &  0.320   & 2.259 &		approx      & GC	\\
\ion{Sr}{i}  &	  7070.07 & -0.030   & 1.847 &		approx      & GC	\\
\ion{Sr}{ii} &	  3380.70 &  0.199   & 2.940 &		approx      & Bh	\\
\ion{Sr}{ii} &	  3464.45 &  0.487   & 3.040 &		approx      & Bh	\\
\ion{Sr}{ii} &	  7334.95 & -1.102   & 7.562 &		approx      & Bh	\\
\hline
\ion{Nd}{ii} &  5130.59  &  0.45    & 1.304	&		-5.61      & HLSC \\
\ion{Nd}{ii} &  5165.13  & -0.74    & 0.680	&		approx      & HLSC \\
\ion{Nd}{ii} &  5255.51  & -0.67    & 0.205	&		approx      & HLSC \\
\ion{Nd}{ii} &  5293.16  &  0.10    & 0.823	&		-5.74      & HLSC \\
\ion{Nd}{ii} &  5319.82  & -0.14    & 0.550	&		-5.82      & HLSC \\
\ion{Nd}{iii}&  5102.16  & -0.62    & 0.296	&		approx      & RRKB \\
\ion{Nd}{iii}&  5294.11  & -0.69    & 0.000	&		approx      & RRKB \\
\ion{Nd}{iii}&  5677.18  & -1.45    & 0.631	&		approx      & RRKB \\
\ion{Nd}{iii}&  5802.53  & -1.71    & 0.296	&		approx      & RRKB \\
\ion{Nd}{iii}&  5845.02  & -1.18    & 0.631	&		approx      & RRKB \\
\ion{Nd}{iii}&  5851.54  & -1.55    & 0.460	&		approx      & RRKB \\
\ion{Nd}{iii}&  6145.07  & -1.33    & 0.296	&		approx      & RRKB \\
\ion{Nd}{iii}&  6327.26  & -1.41    & 0.141	&		approx      & RRKB \\
\ion{Nd}{iii}&  6550.23  & -1.49    & 0.000	&		approx      & RRKB \\
\ion{Nd}{iii}&  6690.83  & -2.46    & 0.460	&		approx      & RRKB \\
\hline
\ion{Pr}{ii} &  4222.90  &  0.271   & 0.055	&		approx      & ILW  \\
\ion{Pr}{ii} &  5322.77  & -0.319   & 0.483	&		approx      & ILW  \\
\ion{Pr}{iii}&  5299.99  & -0.720   & 0.359	&		approx      &ISAN  \\ 
\ion{Pr}{iii}&  6090.01  & -0.871   & 0.359	&		approx      &ISAN  \\
\ion{Pr}{iii}&  6160.23  & -1.020   & 0.173	&		approx      &ISAN  \\
\ion{Pr}{iii}&  6195.62  & -1.071   & 0.000	&		approx      &ISAN  \\
\ion{Pr}{iii}&  7030.39  & -0.929   & 0.359	&		approx      &ISAN  \\
\ion{Pr}{iii}&  7781.98  & -1.276   & 0.000	&		approx      &ISAN  \\
\hline

\footnote{Stark data for Ca-Co are taken from \citet{GFIRON:CaCrFe}. Stark data for rare earth elements are taken from \citet{HLSC}.
Abbreviations for references are as follows:
{\rm Bu}: \citet{Butler1993};
{\rm JK}: \citet{Jonsson1984};
{\rm Wi}: \citet{Wiese1969};
{\rm Ku}: \citet{KuruczCD18};
{\rm $\odot$}: experimental comparison with solar spectrum and HD73666, T. Ryabchikova, L. Fossati (priv. comm.);
{\rm LL}: \citet{LamLuck1978};
{\rm BR}: \citet{BR2002};
{\rm NL}:  \citet{KuruczCD18};
{\rm sH}: solar: Si=7.55 Holweger Model, E. Luck (private communication);
{\rm AJPP}: \citet{AJPP81};
{\rm Wil}: \citet{Wilke03};
{\rm NIST}: \citet{NIST08};
{\rm SG}: \citet{SG1966};
{\rm S}:  \citet{S1988};
{\rm SN}: \citet{SN1975};
{\rm TB}:  \citet{TB1994};
{\rm SR}: \citet{SR1981};
{\rm GF}:  \citet{GFIRON:CaCrFe};
{\rm MFW}: \citet{MFW};
{\rm RU}:  \citet{RaassenUylings1998};
{\rm V2}: various sources - see details in VALD publications \citet{VALD_4,VALD_2,VALD_3};
{\rm FMW}: \citet{FMW};
{\rm SLW}:  \citet{SLW};
{\rm T83av}: average value (Co II) from CUNJ and SLW for the lines with $\log(gf)>-1.0$;
{\rm Bh}: \citet{KuruczCD18};
{\rm GC}: \citet{CarciaCampos1988};
{\rm HLSC}: \citet{HLSC}; 
{\rm RRKB}: \citet{RRKB}; 
{\rm ILW}: \citet{ILW};  
{\rm ISAN}: Ryabtsev, private communication
}

\end{longtable}
} 

\begin{figure*}
\includegraphics[width=0.5\textwidth]{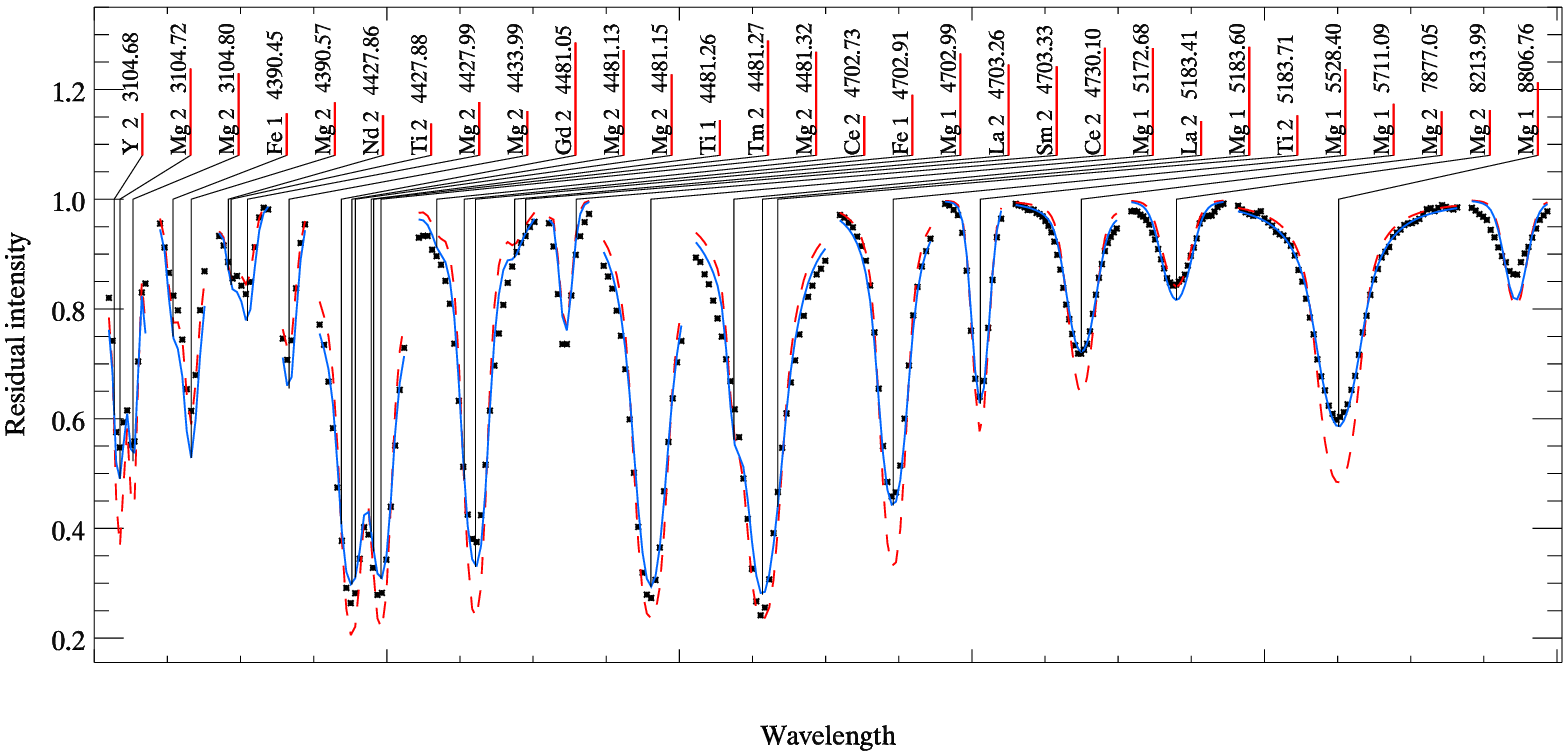}
\includegraphics[width=0.5\textwidth]{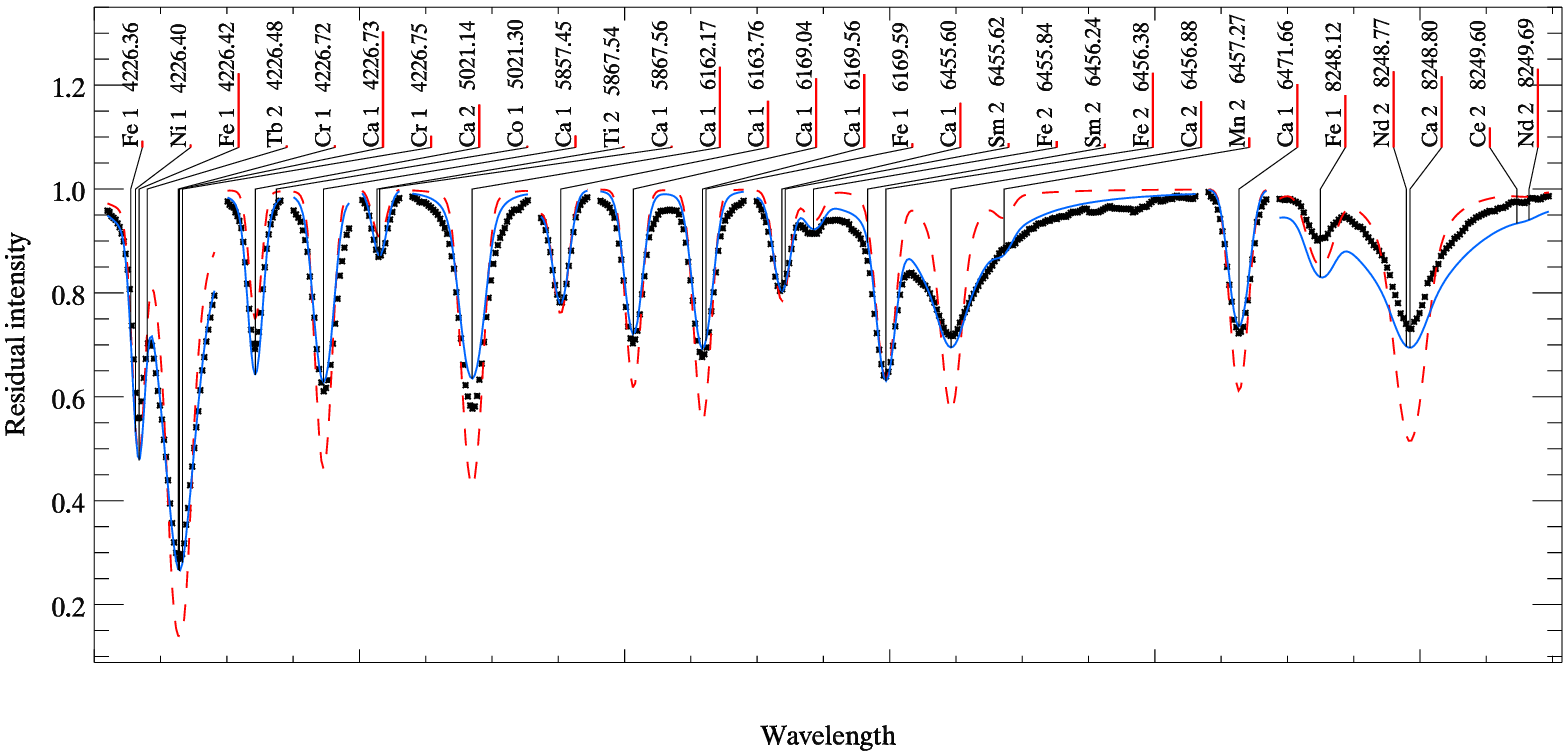}
\caption{Fit of observed (black) and synthetic lines with homogeneous (red) and stratified (blue) abundances. Mg is shown in the left column, Ca in the right column. }
\label{ddafitLines1}
\end{figure*}

\begin{figure*}
\centerline{
\includegraphics[height=0.22\textheight]{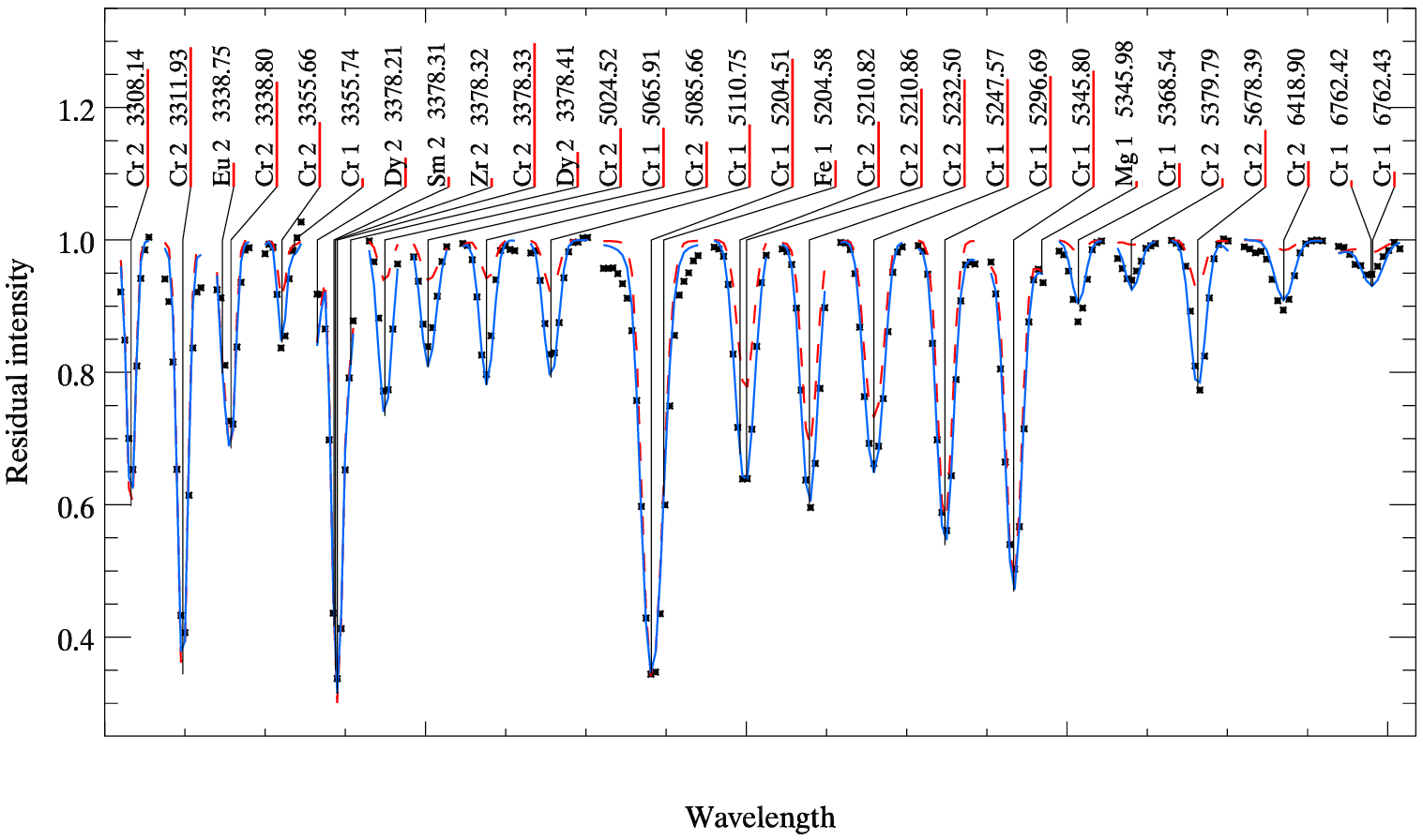}
\includegraphics[height=0.22\textheight]{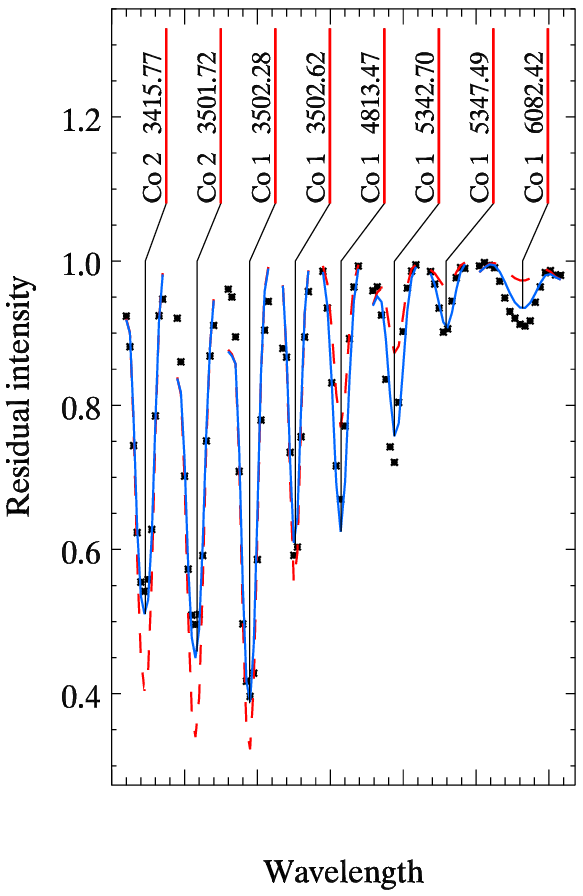}
}
\caption{Fit of observed (black) and synthetic lines with homogeneous (red) and stratified (blue) abundances. Cr is shown in the left column, Co in the right column. }
\label{ddafitLines2}
\end{figure*}

\begin{figure*}
\centerline{
\includegraphics[width=0.5\textwidth]{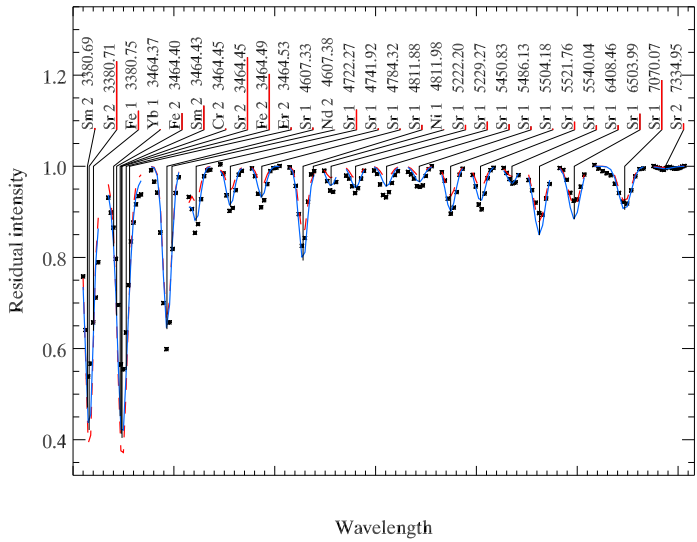}
}
\caption{Fit of observed (black) and synthetic Sr lines with homogeneous (red) and stratified (blue) abundances. }
\label{ddafitLines3}
\end{figure*}

\end{document}